\documentclass[12pt]{article}
\usepackage{amssymb}
\usepackage{amsmath}
\usepackage[utf8]{inputenc}
\usepackage{caption}
\usepackage{booktabs}
\usepackage{subcaption}
\usepackage{multirow}
\usepackage{float} 
\usepackage{adjustbox}
\usepackage{url}

\begin{document}

\title{A Computer Vision--Based Proxy for Political Polarization in Religious Countries: A Türkiye Case Study}
\author{Liangze Ke\thanks{Yale School of Public Health, Yale University; \texttt{liangze.ke@yale.edu}}}
\date{} 
\maketitle

\begin{abstract} 
This paper examines a novel proxy for political polarization, initially proposed by Caliskan et al. \cite{caliskan2025polarization}, which estimates intergroup distances using computer vision. Analyzing 1,400+ YouTube videos with advanced object detection, their study quantifies demographic and religious divides in Türkiye, a deeply polarized nation.

Our findings reveal strong correlations between intergroup distances and electoral polarization, measured via entropy-based voting metrics weighted by religiosity and political inclination. Two key insights emerge: (1) Greater distances between religious and non-religious individuals (NRP vs RP) heighten electoral entropy, underscoring sociocultural fragmentation. (2) Intra-group diversity among non-religious individuals (NRP vs NRP) stabilizes polarization, aligning with Axelrod’s cultural dissemination model.

This research advances computational social science and economics by showing that physical distancing serves as a scalable proxy for polarization, complementing traditional economic indicators.
\end{abstract}


\section{Introduction}
In the evolving landscape of computational social science, the advent of digital technologies has transformed the study of complex social phenomena. Researchers now harness vast datasets from diverse sources, including social media and video platforms, enabling unprecedented insights into human behavior and societal trends. As Lazer et al. \cite{Lazer2009} aptly noted, "We live life in a network," emphasizing the interconnected nature of modern society and the potential of computational approaches to uncover underlying social dynamics. By integrating advanced data processing techniques, this paradigm has opened new avenues for analyzing political polarization, a phenomenon of increasing global relevance.

Conventional social science paradigms have relied on well-established economic indicators, such as Gross Domestic Product (GDP) and income inequality, to quantify the complexity of societal problems. However, these measures face significant criticism for oversimplifying social realities. For instance, GDP fails to account for environmental degradation, unpaid labor, and socio-economic disparities, often leading to a distorted understanding of societal well-being \cite{Kuznets1962}. Similarly, aggregate macroeconomic indicators obscure localized issues and disparities, providing an incomplete picture of polarization across diverse socio-cultural contexts \cite{Alaimo2023}.

Political polarization, a multifaceted social phenomenon, has intensified worldwide, eroding trust in institutions and complicating democratic governance. Economic studies often rely on proxies such as income inequality, government spending, and private-state ownership to explain polarization \cite{Grechyna2016}. However, these variables frequently suffer from endogeneity issues and lack granular coverage, particularly at the district level in developing countries. Moreover, they emphasize economic mechanisms while overlooking critical socio-cultural dimensions, such as religious identity, cultural norms, and racial disparities—factors identified by Alesina et al. \cite{alesina2003fractionalization} and Axelrod's work on cultural dissemination \cite{Axelrod1997} as root causes of societal polarization.

This study proposes an alternative approach: leveraging computer vision techniques to analyze physical walking distances between individuals in public spaces as a proxy for societal divisions and, by extension, political polarization. Walking distances offer a novel method to quantify subtle socio-cultural dynamics, addressing limitations inherent in traditional survey-based approaches. Surveys often suffer from biases, such as the Hawthorne effect, whereas public video footage captures natural human interactions unobtrusively. Walking, as one of the most universal and subconscious forms of interaction, transcends familiarity or unfamiliarity among individuals. Studies such as Enos \cite{enos2017space} and Butler and Brookman \cite{butler2011do} demonstrate that proximity in public spaces can shape attitudes, foster idea dissemination, and reinforce societal boundaries, making walking distances a compelling proxy for socio-political dynamics.

A recent study by Caliskan \cite{caliskan2025polarization} exemplifies this computational approach, using over 1,400 YouTube street videos from Türkiye to measure physical distances between pedestrians. By employing YOLOv5 object detection and 3D relative distance estimation, the study found that shared religious and gender identities significantly reduce interpersonal distances, while the largest gaps occur between non-religious males and religious females. These results reflect entrenched cultural norms and gender boundaries in predominantly Muslim societies, highlighting the potential of computational methods to capture socio-cultural underpinnings of political polarization.

The objective of this paper is to critically evaluate whether walking distances, derived through computer vision techniques, can serve as a robust proxy for political polarization in economic studies. By integrating this innovative approach with traditional economic models, the study aims to bridge computational social science and economics, offering a more nuanced understanding of polarization's socio-economic impacts.

The analysis begins by collecting detailed electoral data, categorizing votes for various parties and candidates, and weighting them based on economic, political, and secularism-related dimensions. Using these weights, entropy values are calculated as proxies for political polarization, reflecting the diversity and alignment of political beliefs expressed through voting behavior. This approach enables linking observable electoral patterns to underlying socio-political divisions.

Secondly, to test the validity of walking distances as a proxy, the study employs Ordinary Least Squares (OLS) regression, incorporating control variables such as regional economic indicators, religiosity, economic sophistication, and environmental factors. Walking distance, the key explanatory variable, is derived from computer vision analysis of public street interactions and represents an innovative method to capture latent social dynamics. The regressions examine whether interpersonal distances between groups—such as religious and non-religious individuals—predict measured electoral polarization, providing insights into the extent to which social distances mirror political divides.

Finally, the paper acknowledges limitations and outlines future research directions. While walking distance offers a compelling proxy for polarization, the study emphasizes the need for benchmarking against existing proxies, such as income inequality or survey-based measures, often unavailable at the provincial level in Türkiye. The paper concludes with a discussion on refining and scaling this methodology, paving the way for further validation in diverse socio-political contexts.

\section{Literature Review}
\label{sec1}

\subsection{Past Studies on Political Polarization}
The concept of political polarization dates back to 1862 when an Englishman lamented “that wretched polarization of our whole national thought… into the two antagonistic currents of common Whiggism and common Toryism” \cite{OED_polarization}. Initially used to describe physical phenomena, the term "polarization" evolved into a widely applied description of political confrontation between competing ideological camps.

Contemporary discourse—journalistic and academic alike—frequently frames polarization as a lamentation of societal dysfunction. For instance, the term "polarization" appears 210 times in just 60 pages of the Democracy Report 2022 by the Varieties of Democracy Institute, which highlights the changing nature of autocratization \cite{Schedler2023}.

Despite extensive work by political scientists and economists, there remains no consensus on the precise definition or causes of political polarization. Existing literature identifies several drivers:

\begin{enumerate}
\item Public and governmental support for wars exacerbates polarization \cite{Lee2022, McCarty2006, Meernik1993, Lee2019}.

\item Religious diversity and conflict contribute significantly to polarization, as evidenced in studies across America \cite{f510ce5e7c7a4ebead2435ddb1bf808b}, Asia \cite{Migheli2019}, and Europe \cite{WilkinsLaflamme2024}.

\item The bifurcation of sexualities and gender identities intensifies societal divisions \cite{Brenner2019, Bragazzi2023}.

\item Racial and ethnic tensions drive polarization, with documented cases globally \cite{Bradley2022}, in the United States \cite{Jardina2022}, and Europe \cite{Evans2002}.

\item Electoral processes and voting outcomes contribute to political divisions \cite{CallanderCarbajal2022, Lachat2008}.

\item The rise of social media amplifies echo chambers and ideological divides on a global scale \cite{Arora2022, Kubin2021} and in specific contexts such as China \cite{Guo2023}.
\end{enumerate}

The impacts of political polarization are also well-documented:

\begin{enumerate}
\item Polarization undermines democratic processes and trust in institutions \cite{Iyengar2012, McCoy2018}.

\item It contributes to policy failures, reducing governmental efficiency \cite{Stanley2020, Fiorina2008}.

\item Polarization erodes social trust and weakens interpersonal cooperation \cite{Reiljan2020, Fiorina2008}.

\item It creates legitimacy barriers, undermining the authority of democratic institutions \cite{Iyengar2012}.

\item Public and personal health outcomes deteriorate in polarized societies \cite{Vasist2023, Fraser2022}.
\end{enumerate}

\subsection{Computer Vision and the Economic Implications of Social Distance}
Political polarization is multidimensional, shaped by both economic and cultural mechanisms. Economists traditionally use proxies like income inequality, government spending, or private-state ownership to measure polarization \cite{Grechyna2016}. However, these measures often fail to capture the cultural and social dimensions underlying divisions.

The concept of social distance, rooted in sociology and social psychology, offers a theoretical lens for understanding physical proximity's reflection of societal fragmentation \cite{Axelrod1997, Heider1946}. Axelrod's Cultural Dissemination framework argues that significant intergroup differences amplify divisions, leading to ideological clustering \cite{Axelrod1997}. Similarly, Heider's Balance Theory emphasizes individuals' efforts to maintain consistency in relationships, reinforcing group boundaries \cite{Heider1946}.

From an economic perspective:

\begin{enumerate}
\item Alesina et al. \cite{alesina2003fractionalization} show that social and cultural divisions, including ethnic and religious fractionalization, reduce economic performance by lowering public goods provision, increasing corruption, and weakening institutional effectiveness.

\item Barro \cite{Barro2001human} demonstrates how cultural heterogeneity hinders trust and cooperation, impacting democratic efficiency and economic growth.
\end{enumerate}

Unlike survey-based measures prone to biases, physical walking distances provide an objective proxy for social distance.

\begin{enumerate}
\item Dietrich and Sands \cite{DietrichSands2023} use traffic camera footage to analyze racial dynamics in New York City, linking physical distancing to intergroup tensions.

\item Enos \cite{enos2017space} highlights that proximity in public spaces predicts intergroup attitudes and trust.

\item Caliskan et al. \cite{caliskan2025polarization} measure interpersonal distances in Türkiye using YOLOv5 object detection, finding that shared religious and gender identities reduce distances, while non-religious males and religious females exhibit the largest gaps.
\end{enumerate}

This study extends their findings to an economic framework, demonstrating the correlation between walking distances and electoral polarization, bridging computational social science with economic theory.

\subsection{Türkiye and Beyond}

In recent years, the intensifying political polarization between secularism and political Islam within Turkish society has garnered substantial scholarly attention. A common street scene in cities like Ankara or Istanbul reflects this divide: a woman in elegant, fashionable attire walks past another woman, a devout believer, adorned in a vibrant hijab—a cultural staple in Turkey— and draped in heavy, silky fabric that covers most of her head and body. This juxtaposition highlights the polarized state of Turkish streets. Under- standing the scale and implications of this polarization is critical for devising strategies to mitigate its adverse effects and comprehending its political con- sequences. This project adopts a multidisciplinary approach, combining ad- vanced computer vision techniques with statistical methodologies to explore the potential causal links between identity and political polarization.

The headscarf, known as the hijab in some other cultures, has been a crucial cultural identifier and, in some instances, a significant political symbol in the lives of religious women in Turkey. It represents an embodiment of their pious beliefs and Islamic identity \cite{slininger2014}. Political Islam plays a pivotal role in Turkey’s politics and serves as a driving force behind polarization. “There is significant tension around the issue of secularism or laicism in the country” \cite{carkoglu2007}

Turkiye serves as a compelling case study for understanding political po- larization due to its stark ideological divides, religious dynamics, and socio- political history. These characteristics not only offer a unique setting for studying polarization but also provide an opportunity to test the replicabil- ity of computational methodologies in analogous contexts. Similar dynam- ics can be observed in other religiously influenced nations, such as Iran or Israel, where public behaviors and attire serve as visible markers of ideolog- ical affiliations. By leveraging the Turkish case, this study aims to refine methodologies that can be applied to other polarized societies, contributing to a broader understanding of how identity shapes political behavior.

There is strong replicability of its methods in societies with comparable polarization dynamics, such as Iran or Israel, where visible markers of cultural or religious identity play a central role in public life. In Iran, the strict enforcement of Islamic dress codes creates clear visual distinctions between social groups, making it relatively straightforward to apply the methodologies developed for Türkiye to study socio-political divides. Similarly, in Israel, the ideological cleavages between secular Jews, religious Jews, and Arabs are often reflected in public behaviors and attire, providing another suitable context for replication. The visibility of these markers facilitates the use of advanced computer vision techniques to analyze polarization, demonstrating the adaptability of this approach. By leveraging the case of Türkiye, this study highlights the potential for these methods to generate insights in diverse cultural and political environments, emphasizing their global relevance.

\section{Data}
This section describes the dataset used in the paper to explore the validity of the proxy variable - the relative distances extracted using the CV approach. The dataset is merged by two parts - the election dataset, and the CV relative distance dataset. It will be referred to as "election data" and "distance data" accordingly in this paper. 
\subsection{Election Data}
The election data sources from \cite{turkeyelections}. Two datasets are involved, which are documented as "\textit{cb\_2018\_il}" (2018 presidential election) and "\textit{yerel\_secim\_2019\_il}" (2019 local election).
Both datasets are at the provincial level, meaning each dataset records voting information across all 81 provinces in Turkiye. 2018 presidential election dataset involves 105 variables, which could be summarized into 3 categories: 1)the unit of observation: 81 provinces, 2) general votes (registered voters, turnouts, and valid votes), 3) candidates votes (including votes and share of votes obtained by presidential candidates). Similarly, the 2019 local election dataset contains 31 variables, including 1) the unit of observation: 81 provinces, 2) general votes (registered voters, turnouts, and valid votes), and 3) party votes(including votes and share of votes obtained by each party).

\subsection{Distance Data}
This distance dataset is documented by \cite{caliskan2025polarization}. It recorded 16833 pieces of pictures, and the relative distance information accordingly. It randomly samples from 81 provinces with slightly uneven patterns. It contains 24 variables, including 1) metadata of sources (including video ID, framing information, video publish information), 2) photo location (spanning 81 provinces), and 3) distances (including NRP\_vs\_NRP: distance between non-religious people and non-religious-people,	RP\_vs\_RP: distance between religious people and religious people,	NRP\_vs\_RP: non-religious people and religious people.)

\begin{figure}[ht]
    \centering
    \begin{minipage}[b]{0.48\textwidth}
        \centering
        \includegraphics[width=\linewidth]{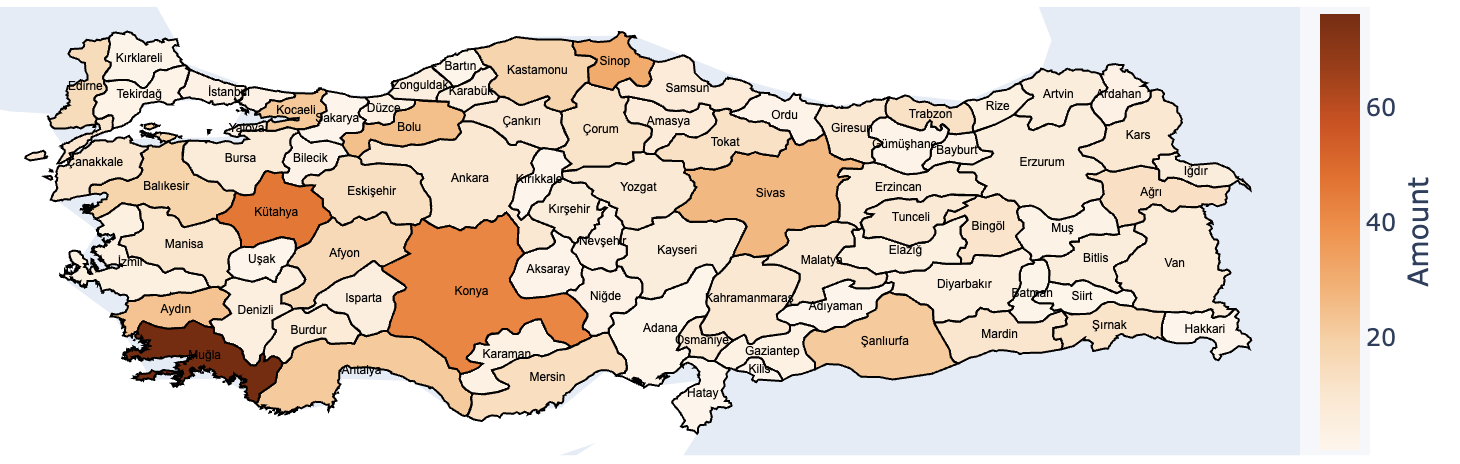}
        \caption{The Count of Religious People (RP) in Distance Data\cite{caliskan2025polarization}}
        \label{fig:religious-map}
    \end{minipage}
    \hfill
    \begin{minipage}[b]{0.48\textwidth}
        \centering
        \includegraphics[width=\linewidth]{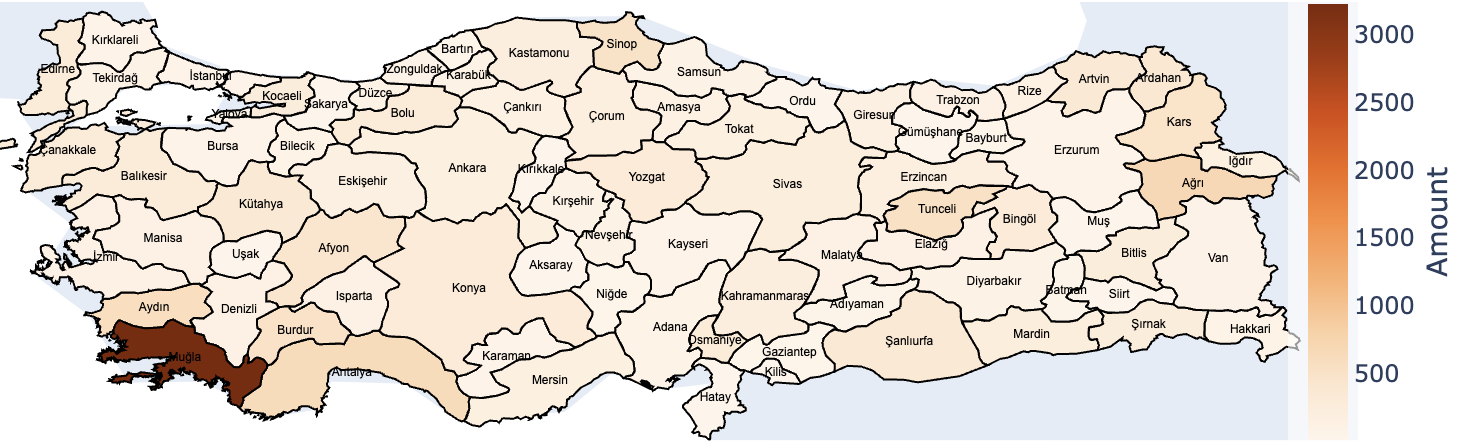}
        \caption{The Count of Non-Religious People (NRP) in Distance Data\cite{caliskan2025polarization}}
        \label{fig:nonreligious-map}
    \end{minipage}
\end{figure}

\subsection{Merged Dataset and Processing}
The final dataset includes voting entropy measures appended to 81 distinct provinces in Türkiye. In addition to these entropies, key control variables have been included to provide a comprehensive framework for analysis. These controls encompass per capita GDP at the provincial level and contributions to GDP from manufacturing, services, agriculture, and finance/insurance sectors. These variables aim to capture variations in provincial wealth, industrial structure, and economic sophistication.

By combining voting entropies and these controls, the dataset enables robust analysis of political and economic patterns across Türkiye. Each province serves as a unique observational unit, offering granularity and diversity for empirical analysis.

The control variables were sourced from the Turkish Statistical Institute (TurkStat) \cite{turkstat} to account for geographic and economic disparities across provinces. GDP per capita for each province, along with GDP contributions across economic sectors, were appended to the merged dataset. Provincial GDP contributions were divided into several economic channels to indicate the development and industrial specialization of each province:

\begin{enumerate}
\item \textit{Manufacturing GDP:} Represents the share of provincial GDP derived from manufacturing activities, reflecting the industrial base of the region.

\item \textit{Service GDP:} Captures contributions from the service sector, including retail, tourism, and professional services, which often dominate in more urbanized and developed regions.

\item \textit{Agriculture GDP:} Reflects the share of GDP from agriculture, forestry, and fishing. This variable highlights the rural economic base and the reliance on traditional sectors in certain provinces.

\item \textit{Financial and Insurance Activities GDP:} Represents the contributions of financial services and insurance, serving as a key indicator of economic sophistication and urbanization.
\end{enumerate}

These channels provide critical information on the status of development and the relative poverty or wealth of each province. Provinces with higher shares of manufacturing and financial activities GDP are likely to be more developed and economically sophisticated, whereas those reliant on agriculture tend to exhibit lower levels of economic development and sophistication.

\begin{figure}[ht]
    \centering
    \begin{minipage}[b]{0.45\textwidth}
        \centering
        \includegraphics[width=\linewidth]{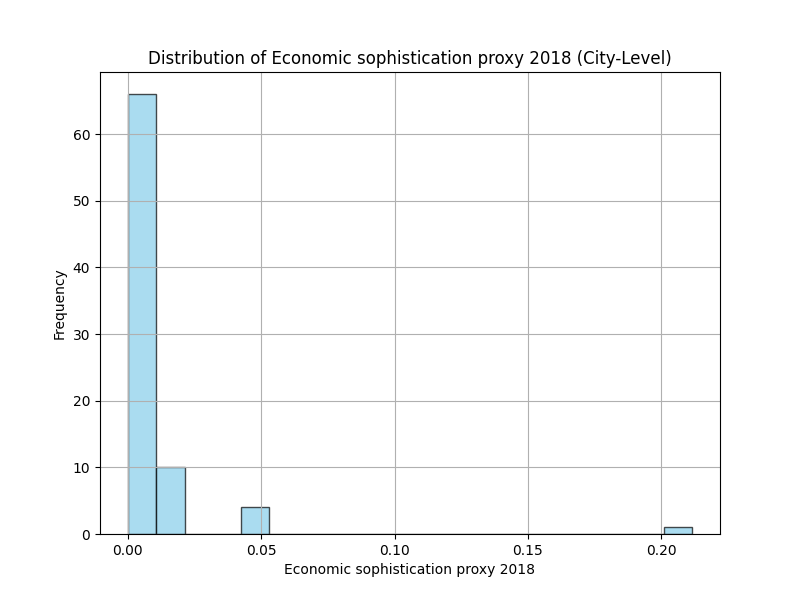}
        \caption{Economic Sophistication Proxy (2018)}
        \label{fig:economic_sophistication_2018}
    \end{minipage}
    \hfill
    \begin{minipage}[b]{0.45\textwidth}
        \centering
        \includegraphics[width=\linewidth]{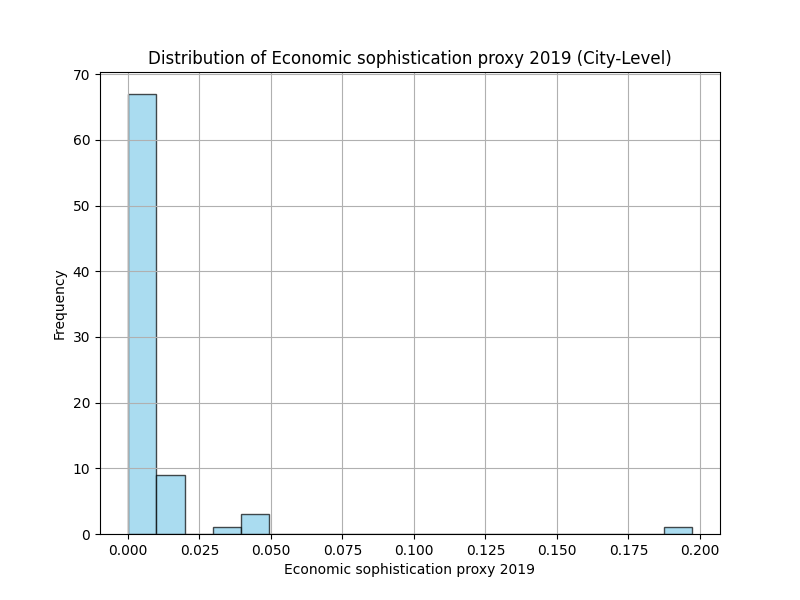}
        \caption{Economic Sophistication Proxy (2019)}
        \label{fig:economic_sophistication_2019}
    \end{minipage}
    \vspace{1em}
    \begin{minipage}[b]{0.45\textwidth}
        \centering
        \includegraphics[width=\linewidth]{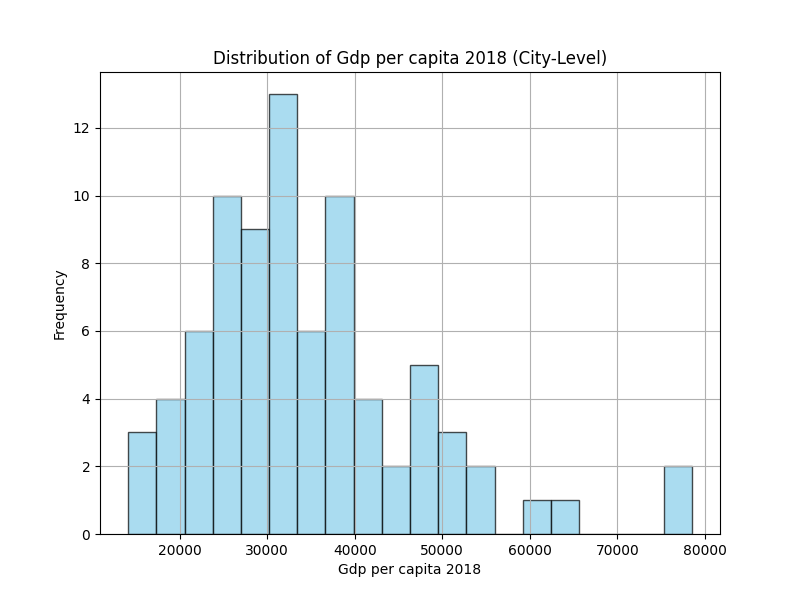}
        \caption{GDP per Capita (2018)}
        \label{fig:gdp_per_capita_2018}
    \end{minipage}
    \hfill
    \begin{minipage}[b]{0.45\textwidth}
        \centering
        \includegraphics[width=\linewidth]{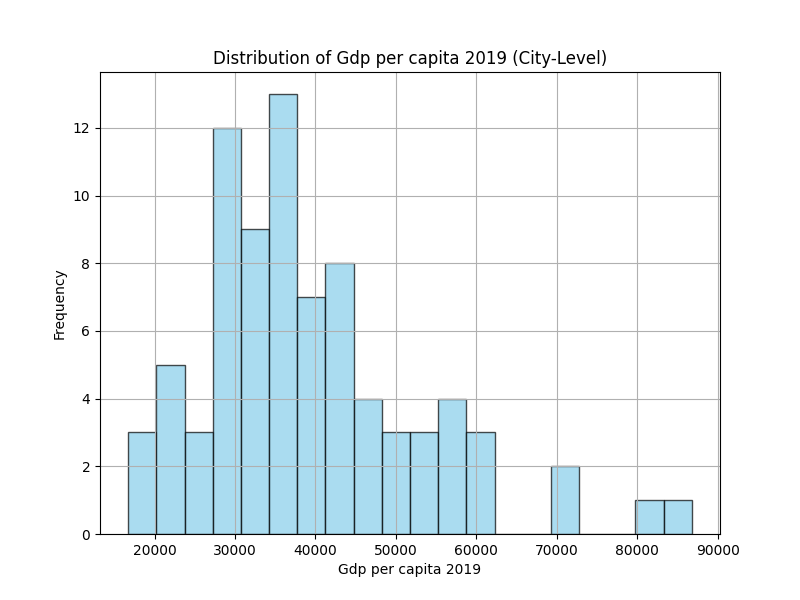}
        \caption{GDP per Capita (2019)}
        \label{fig:gdp_per_capita_2019}
    \end{minipage}
    \caption{Distributions of Economic Sophistication Proxy and GDP per Capita by Year}
    \label{fig:control_distribution}
\end{figure}

The distribution of GDP per capita at the provincial level for 2018 and 2019 (see Figure~\ref{fig:control_distribution}) reveals significant regional economic disparities across Türkiye. The histograms for both years exhibit a right-skewed distribution, indicating that a majority of provinces have relatively low GDP per capita. In contrast, a small number of urbanized provinces, such as Istanbul and Ankara, display substantially higher values. This observation underscores the uneven economic development within the country, with wealth concentrated in a few urban regions while economic challenges persist in many rural provinces.

\section{LHS: Proxies for Political Polarization}

\subsection{Shannon Entropy and Rethinking About Polarization}

One straightforward way to assess polarization is by examining the spread of opinions, essentially how varied and scattered public opinions are. DiMaggio, Evans, and Bryson refer to this as "dispersion," describing it as the condition in which opinions differ significantly and are widely separated in content \cite{DiMaggio1996}. They also propose a principle of dispersion: "All else being equal, the greater the dispersion of opinions, the harder it will be for the political system to create and sustain a centrist political consensus".

People’s socio-political views can be reflected through voting systems, as candidates represent the spectrum of voters' political preferences \cite{downs1957}.

Shannon entropy \cite{shannon1948} quantifies the uncertainty or information content in a probability distribution, measuring the expected value of the information contained in a random variable's outcomes. Since Turkiye is a multi-party nation, Shannon entropy is useful for measuring how "diverse" every election is.

\begin{equation}
H = - \sum_{i=1}^{n} p_i \log p_i
\label{eq:entropy}
\end{equation}

\text{where:}

\begin{itemize}
    \item $H$: \text{Entropy, which measures the uncertainty or diversity of the vote distribution.}
    \item $p_i$: \text{The proportion of votes received by party $i$.}
    \item $n$: \text{The total number of political parties.}
\end{itemize}

\cite{turkeyelections} publicize the election results of Turkiye from 1995 to 2023. (The details of the dataset are described in \textit{Data} section) The 2018 Presidential Election and the 2019 Local Election in Türkiye were selected as indicators of voters' political preferences. \textit{2018\_presidential\_entropy} and \textit{2019\_local\_entropy} were calculated as proxies for the diversity of voter's opinions on a provincial basis. 
\begin{figure}
    \centering
    \includegraphics[width=\linewidth]{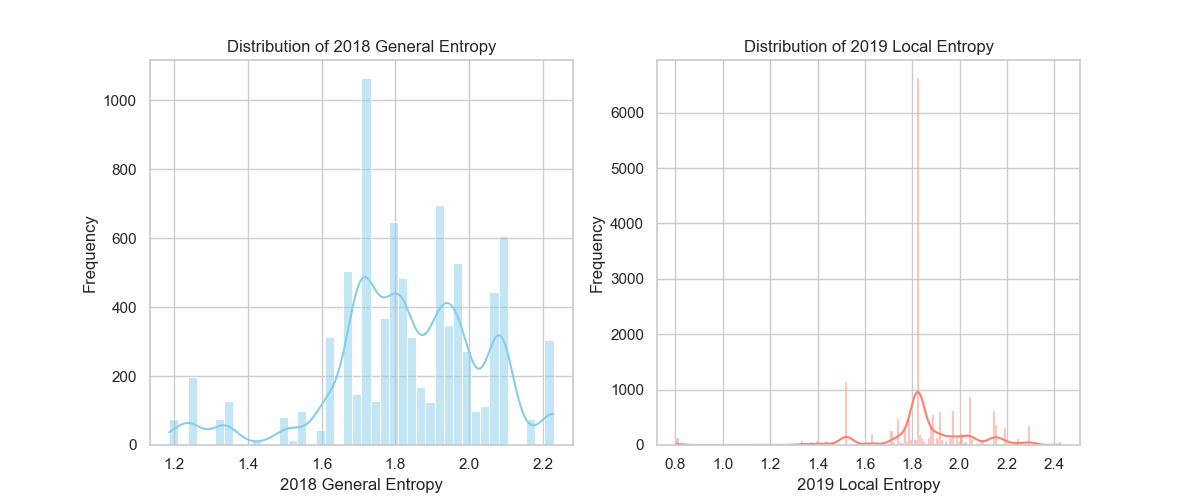}
    \caption{Entropy Distributions}
    \label{fig:entropy}
\end{figure}

\begin{figure}[h!]
    \centering
    \begin{subfigure}[b]{0.75\textwidth}
        \includegraphics[width=\textwidth]{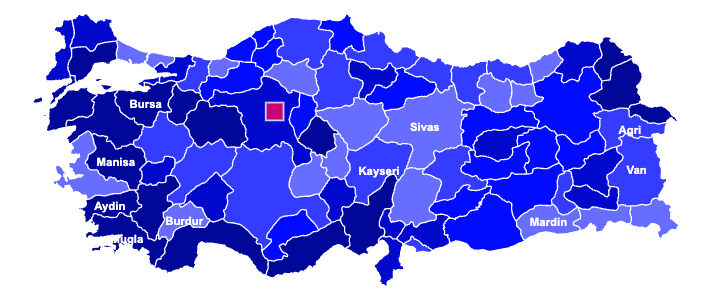}
        \caption{2018 General Election Entropy Value Map}
        \label{fig:2018en}
    \end{subfigure}
    
    \vspace{1em} 

    \begin{subfigure}[b]{0.75\textwidth}
        \includegraphics[width=\textwidth]{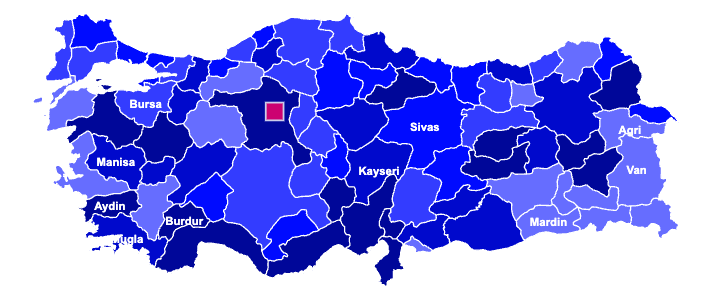}
        \caption{2019 Local Election Entropy Value Map}
        \label{fig:2019en}
    \end{subfigure}

    \caption{Entropy Value Maps for Turkish Elections: 2018 General Election and 2019 Local Election}
    \label{fig:entropy_maps}
\end{figure}

As figure \ref{fig:2018en} and figure \ref{fig:2019en}\ref{fig:entropy} show: the analysis of entropy distributions for the 2018 General Election and the 2019 Local Election across Turkish provinces reveals a shift towards greater polarization in voting behavior. In 2018, entropy values were more dispersed, with a moderate level of diversity across provinces, reflecting a balanced heterogeneity in voter preferences. However, by 2019, entropy values peaked sharply around 1.8, indicating a highly polarized state. This shift likely corresponds to the aftermath of the 2017 constitutional referendum, which granted President Erdoğan expanded executive powers. The resulting centralization may have intensified partisan divides and consolidated political loyalties, driving a more uniform, polarized electoral landscape in the 2019 local elections.

\subsection{Weighting the Polars: Upgraded Shannon Entropy}
Equation \ref{eq:entropy} treats every political party or candidate with equal value. However, not every party or candidate has an equal impact on social polarization. Political spectrum is often used in political science to summarize a political ideology's position in terms of some major aspects\cite{downs1957}. 

For the two election datasets, the 2018 presidential election and the 2019 local election, each candidate or party is summarized based on four perspectives: Nationalism, Secularism, Democracy \& Governance, and Economics Policy. Depending on the ideologies summarized, a religiosity degree and a political spectrum degree are appended. These serve as weights that will be updated to the equation of entropy \ref{eq:entropy}. The 2018 presidential election is summarized in Table\ref{table:candidates_2018_part1}, and the 2019 local election is summarized in Table \ref{table:parties_2019_part1}.

\begin{table}[h]
\centering
\footnotesize
\begin{tabular}{|p{2cm}|p{3cm}|p{4cm}|p{4cm}|}
\hline
\textbf{Candidate} & \textbf{Party} & \textbf{Nationalism} & \textbf{Secularism} \\ \hline
Muharrem İnce & Republican People's Party (CHP) & Moderate nationalism, supports EU membership \cite{hurriyet_ince_nomination} & Strong secularism, opposes religious influence \cite{sozcuk_kesici} \\ \hline
Meral Akşener & İyi Party & Strong nationalism, emphasizes Turkish unity \cite{turkish_minute_aksener_signatures} & Supports secularism \cite{turkish_minute_aksener_signatures} \\ \hline
Recep Tayyip Erdoğan & Justice and Development Party (AKP) & Assertive nationalism, neo-Ottoman influence \cite{hurriyet_erdogan_nomination} & Limited secularism, conservative policies \cite{hurriyet_erdogan_nomination} \\ \hline
Selahattin Demirtaş & Peoples' Democratic Party (HDP) & Advocates for minority rights, Kurdish autonomy \cite{hurriyet_demirtas_nomination} & Strong secularism, promotes pluralism \cite{hurriyet_demirtas_nomination} \\ \hline
Temel Karamollaoğlu & Felicity Party (SP) & National sovereignty, conservative approach \cite{eralp_turkey_candidates_eu} & Supports ethical values from Islam \cite{karamollaoglu_socialmedia} \\ \hline
Doğu Perinçek & Patriotic Party (Vatan Partisi) & Strong nationalism, anti-foreign intervention \cite{aydinlik_perincek_signatures} & Strong secularism, Kemalist values \cite{aydinlik_perincek_signatures} \\ \hline
\end{tabular}
\caption{Summary of Turkish 2018 Presidential Candidates' Political Positions (Part 1)}
\label{table:candidates_2018_part1}
\end{table}

\clearpage
\begin{table}[H]
\ContinuedFloat
\centering
\footnotesize
\begin{tabular}{|p{2cm}|p{3cm}|p{3cm}|p{2cm}|p{2cm}|}
\hline
\textbf{Candidate} & \textbf{Democracy \& Governance} & \textbf{Economic Policy} & \textbf{Religiosity Degree (0-1)} & \textbf{Political Spectrum (0-1)} \\ \hline
Muharrem İnce & Advocates for democratic norms, rule of law \cite{hurriyet_ince_nomination} & Focuses on sustainable economic reforms \cite{hurriyet_ince_president} & 0.2 & 0.3 \\ \hline
Meral Akşener & Advocates for checks and balances, transparency \cite{turkish_minute_aksener_signatures} & Focuses on economic stability \cite{turkish_minute_aksener_signatures} & 0.4 & 0.6 \\ \hline
Recep Tayyip Erdoğan & Centralized governance \cite{balkan_bosnia_rally} & Emphasizes rapid development \cite{hurriyet_joint_application} & 0.8 & 0.8 \\ \hline
Selahattin Demirtaş & Advocates for democratic reforms \cite{hurriyet_demirtas_nomination} & Focuses on social welfare \cite{hurriyet_demirtas_nomination} & 0.2 & 0.2 \\ \hline
Temel Karamollaoğlu & Advocates democracy with Islamic ethics \cite{karamollaoglu_socialmedia} & Social justice, equitable wealth distribution \cite{karamollaoglu_socialmedia} & 0.7 & 0.7 \\ \hline
Doğu Perinçek & State sovereignty, centralized governance \cite{aydinlik_perincek_signatures} & State-led economy, industrialization \cite{aydinlik_perincek_signatures} & 0.3 & 0.3 \\ \hline
\end{tabular}
\caption{Summary of Turkish 2018 Presidential Candidates' Political Positions (Part 2)}
\label{table:candidates_2018_part2}
\end{table}

\clearpage
\begin{table}[H]
\centering
\footnotesize
\begin{tabular}{|p{4cm}|p{5cm}|p{5cm}|}
\hline
\textbf{Party} & \textbf{Nationalism} & \textbf{Secularism} \\ \hline
SAADET (Felicity Party) & Emphasizes Turkish nationalism with a focus on Islamic values.\cite{saadet} & Supports secularism but advocates for greater public visibility of religion. \\ \hline
BTP (Independent Turkey Party) & Advocates for Turkish nationalism and independence.\cite{btp} & Supports secularism with respect for religious values. \\ \hline
TKP (Communist Party of Turkey) & Opposes nationalism, promoting internationalism and class solidarity. & Strongly supports secularism and opposes religious influence in politics. \\ \hline
VP (Patriotic Party) & Strong emphasis on Turkish nationalism and sovereignty. & Supports strict secularism in line with Kemalist principles. \\ \hline
BBP (Great Unity Party) & Combines Turkish nationalism with Islamic values. & Supports secularism but with a significant role for religion in public life. \\ \hline
CHP (Republican People's Party) & Emphasizes civic nationalism based on citizenship.\cite{chp} & Strongly supports secularism as a core principle. \\ \hline
AKP (Justice and Development Party) & Emphasizes Turkish nationalism with a conservative outlook. & Supports secularism but has increased the visibility of religion in public life. \\ \hline
DP (Democrat Party) & Emphasizes Turkish nationalism with a liberal-conservative approach. & Supports secularism with respect for religious freedoms. \\ \hline
MHP (Nationalist Movement Party) & Strong emphasis on Turkish nationalism and unity.\cite{mhp} & Supports secularism but with respect for traditional values. \cite{baskan2006}\cite{celep2010} \\ \hline
IYI (Good Party) & Emphasizes Turkish nationalism with a centrist approach.\cite{kellam2017}\cite{luca2023} & Supports secularism and the separation of religion and state.\cite{iyi} \\ \hline
HDP (Peoples' Democratic Party) & Advocates for minority rights and multiculturalism. & Supports secularism and opposes religious influence in politics. \\ \hline
DSP (Democratic Left Party) & Emphasizes social democracy with a focus on social justice. & Supports secularism as a fundamental principle. \\ \hline
\end{tabular}
\caption{Summary of Turkish Political Party Positions in 2019 (Part 1)}
\label{table:parties_2019_part1}
\end{table}

\begin{table}[H]
\ContinuedFloat
\centering
\footnotesize
\begin{tabular}{|p{2.5cm}|p{4cm}|p{4cm}|p{2cm}|p{2cm}|}
\hline
\textbf{Party} & \textbf{Governance \& Democracy} & \textbf{Economic Policy} & \textbf{Religiosity Degree (0-1)} & \textbf{Political Spectrum (0-1)} \\ \hline
SAADET (Felicity Party) & Promotes democratic governance with an emphasis on moral and ethical values.\cite{saadet} & Advocates for a mixed economy with social justice principles\cite{saadet2011}. & 0.8 & 0.3 \\ \hline
BTP (Independent Turkey Party) & Supports democratic principles with a focus on national sovereignty. & Proposes a unique economic model emphasizing national resources. & 0.6 & 0.4 \\ \hline
TKP (Communist Party of Turkey) & Advocates for a socialist state with centralized planning.\cite{tkp} & Supports state-controlled economy and wealth redistribution. & 0.1 & 0.1 \\ \hline
VP (Patriotic Party) & Advocates for a strong state with a focus on national unity.\cite{vp} & Supports state intervention in the economy and protectionism. & 0.2 & 0.2 \\ \hline
BBP (Great Unity Party) & Advocates for democratic governance with a focus on national and moral values.\cite{bbp} & Supports a free-market economy with social justice considerations. & 0.7 & 0.3 \\ \hline
CHP (Republican People's Party) & Advocates for parliamentary democracy and rule of law.\cite{chp} & Supports a mixed economy with social democratic policies\cite{tavits2008} & 0.2 & 0.6 \\ \hline
AKP (Justice and Development Party) & Advocates for democratic governance with a strong executive branch.\cite{AKP} & Supports a free-market economy with neoliberal policies.\cite{gumuscu2024}\cite{yesilada2023} & 0.6 & 0.7 \\ \hline
DP (Democrat Party) & Advocates for democratic governance with an emphasis on individual rights.\cite{dp} & Supports a free-market economy with liberal policies. & 0.5 & 0.5 \\ \hline
MHP (Nationalist Movement Party) & Advocates for a strong state with a focus on national security.\cite{mhp} & Supports a mixed economy with state intervention when necessary.\cite{kellam2017}\cite{kidal2020} & 0.4 & 0.3 \\ \hline
IYI (Good Party) & Advocates for democratic governance and rule of law.\cite{iyi} & Supports a free-market economy with social policies.\cite{cengiz2021} & 0.3 & 0.5 \\ \hline
HDP (Peoples' Democratic Party) & Advocates for democratic governance with a focus on human rights. \cite{hdp}& Supports a mixed economy with social justice policies.\cite{secen2024} & 0.2 & 0.4 \\ \hline
DSP (Democratic Left Party) & Advocates for democratic governance and social welfare.\cite{dsp} & Supports a mixed economy with state intervention for social equity. & 0.2 & 0.4 \\ \hline
\end{tabular}
\caption{Summary of Turkish Political Party Positions in 2019 (Part 2)}
\label{table:parties_2019_part2}
\end{table}
To measure the diversity of political preferences within each district, we use an entropy formula that incorporates each party’s religiosity (\( r_i \)), political spectrum position (\( s_i \)), and vote percentage (\( p_i \)).  

Let \( p_i \) be the percentage of votes received by each party in a district, \( r_i \) the religiosity level of each party, and \( s_i \) the political spectrum position of each party, where lower values indicate left-wing and higher values indicate right-wing positions.  

The entropy \( H \) for each district, representing political diversity based on these combined characteristics, is calculated as follows:  

\begin{equation}  
H(religiosity) = - \sum_{i=1}^{n} p_i \cdot r_i \cdot \log(p_i \cdot r_i)  
\label{eq:entropy_r}  
\end{equation}  

\begin{equation}  
H(political) = - \sum_{i=1}^{n} p_i \cdot s_i \cdot \log(p_i \cdot s_i)  
\label{eq:entropy_p}  
\end{equation}  

Two updated entropy values are calculated alongside the original entropy values, based on the 2018 presidential election data and the 2019 local election data. The updated entropies are referred to as:  

\begin{enumerate}  
    \item Weighted by Religiosity Degree: \( H(religiosity) \)  
    \item Weighted by Political Spectrum Degree: \( H(political) \)  
\end{enumerate}  

This approach provides flexibility in adjusting the diversity measure to reflect local political characteristics. By incorporating religiosity and ideological stance into the entropy calculation, the resulting \( H \) captures both the cultural and ideological dimensions of vote distributions, offering a more nuanced understanding of political diversity.

\section{RHS: Physical Distance Between Pedestrians on Turkiye Street}

\subsection{Distance variables}
This paper\cite{caliskan2025polarization} is motivated by the growing political polarization globally, particularly the ideological divide between secularism and political Islam in Turkey. The paper analyzes over 1,400 publicly available YouTube videos filmed on the streets of Turkey, extracting approximately 170,000 frames featuring pedestrians. Using the YOLOv5 algorithm, authors detect and classify pedestrians based on gender and religiosity. Additionally, the authors develop and refine two novel distance estimation methods to measure the relative distances between pairs of pedestrians. Their innovative approach enables the conversion of 2D distances in street videos into 3D relative distances between pedestrians with varying genders and religiosity levels.

Technically, the authors of the paper scraped randomly from YouTube, downloading 1400 Turkey street videos that meet their criteria. 170,000 frames were randomly extracted from the videos, and the relative distances between pedestrians of different religiosity in the pictures were recorded. The output dataset cited from \cite{caliskan2025polarization} is detailed in section \textit{Data}. For technical details, the reader shall refer to \cite{caliskan2025polarization} and Appendix A below.

\subsection{Control Variables}
Three main categories of control variables are involved in this paper: 1. regional religiosity, 2. regional degree of economic sophistication, and 3. environment. Worth mentioning, time and region fix effects are accounted for, but are not expanded in this section.

First, to account for the level of religiosity across provinces, the variable \textit{\#temples} is used as a control. This variable estimates the number of religious structures within each province. It was derived using a geospatial Python script that utilized the Google Places API to search for terms such as "mosques" and "temples," and the total number of search results was recorded.  

The numbers of registered Muslim population in Türkiye provinces are not used.  Becasue this could be misleading. Despite Turkey's official record of 99\% of the population being registered as Muslim\cite{us_religious_freedom_turkey_2021}, studies have suggested that a significant proportion of these individuals do not actively adhere to Islamic beliefs or practices. Many are secular in lifestyle, with minimal engagement in religious culture or rituals. Registering for Muslim officially is based of other motivations, including gaining a political advantage. Often, those registered Muslims behaving a secular lifestyle often having political opinions for secularism or even Kemalism, maintaining that the religion should not be extended to political influence. Those people are often referred to as "Islamic-secular" by scholars. \cite{ackson2017islamicsecular} 

Thus, using \textit{\#temples} as a proxy for religiosity circumvents the limitations of relying on self-reported religious affiliation, which may not accurately reflect the cultural and religious norms in practice.
The inclusion of this control addresses potential biases arising from varying perceptions of religious and non-religious individuals across provinces. Regions with differing levels of religiosity may interpret the "distance" between religious and non-religious populations differently. This could affect voting patterns or political entropy.

\begin{figure}
    \centering
    \includegraphics[width=0.75\linewidth]{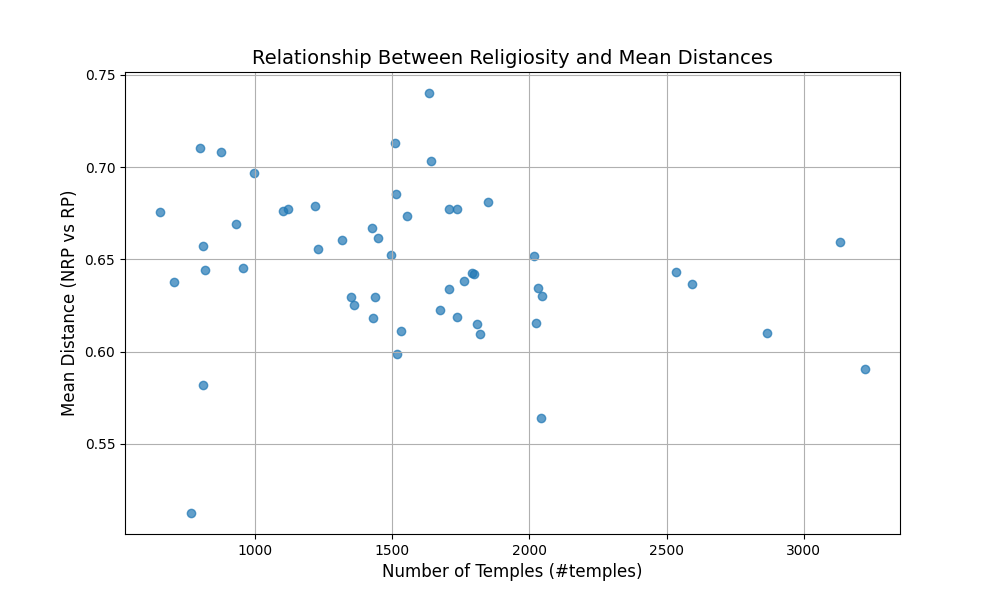}
    \caption{Relationship between number of temples and intersocial walking distance}
    \label{fig:relationship_religiosity_distance}
\end{figure}

To evaluate the relationship between religiosity (proxied by \textit{\#temples}) and the perceived "distance" between political or cultural groups, the mean distance (\textit{NRP vs RP}) against \textit{\#temples} is plotted. The scatterplot (Figure \ref{fig:relationship_religiosity_distance}) demonstrates no strong correlation between these variables, suggesting that religiosity does not significantly drive differences in perceived distances across regions. This implies a consistent perception of "distance" irrespective of religiosity.

Second, to account for the economic conditions of each province, the dataset includes an economic sophistication proxy. This proxy reflects the weight of manufacturing and industrial activities in the economy of each province. Specifically, the proxy for each year is calculated as follows:

The economic proportions for agriculture, industry, and services, along with the Economic Sophistication Proxy, are calculated as follows:

\begin{align}
    \text{Agriculture Proportion} &= \frac{\text{Agriculture GDP}}{\text{Total GDP}} \\
    \text{Industry Proportion} &= \frac{\text{Manufacturing GDP} + \text{Industry GDP}}{\text{Total GDP}} \\
    \text{Services Proportion} &= \frac{\text{Services GDP} + \text{Finance/Insurance GDP}}{\text{Total GDP}} \\
    \text{Economic Sophistication Proxy} &= \frac{\text{Manufacturing GDP} + \text{Industry GDP}}{\text{Total GDP}}
\end{align}

The Economic Sophistication Proxy primarily reflects the degree of industrial and productive development in each province. Industrialization and manufacturing are considered key indicators of economic complexity and development \cite{hausmann2007complexity}. 

The sectoral proportions for agriculture, services, and finance/insurance provide additional context regarding each province's economic structure: for instance, a high agricultural proportion indicates a reliance on low-value-added activities, which are typically associated with rural and less developed regions. On the other hand, a high proportion of services and financial indicates a relatively sophisticated region. The underlying assumption is that provinces with higher shares of industrial and manufacturing activities tend to be more economically developed and sophisticated. 

The Economic Sophistication Proxy addresses potential confounders by ensuring that variations in voting entropy are not merely artifacts of economic differences across regions. For example, provinces with underdeveloped economies may exhibit different patterns of political behavior compared to highly industrialized ones. By controlling for economic sophistication, we aim to isolate the effects of other covariates, such as religiosity and urbanization, on voting entropy.

Third, environmental controls are also attached. The dataset contains observations extracted from YouTube videos, where frames capture individuals in various settings. In colder seasons, individuals are more likely to wear heavy winter clothing, such as scarves, leather jackets, and coats. These garments may obscure religious symbols (e.g., headscarves, robes), leading to the underrepresentation of visible religiosity in certain frames. Conversely, during summer months, lighter clothing typically provides better visibility of religious symbols, allowing the system to detect and classify attire more accurately. To account for environmental factors that may impact the analysis, we include the following binary variables:
\begin{enumerate}
    \item daynight: A binary variable indicating whether the footage was taken during the day (1) or at night (0).
    \item is\_summer: A binary variable indicating whether the footage was recorded during warmer months (1) or colder months (0).
\end{enumerate}

\section{OLS Regression and Mechanism}
The relationship between political polarization and the selected control variables is captured using the following regression model:

\begin{equation}
\begin{aligned}
\text{Political Polarization}^t &= \beta_0 
+ \beta_1 \text{NRP vs RP}^t 
+ \beta_2 \text{NRP vs NRP}^t \\
&\quad + \gamma_1 \text{num\_mosques}^t 
+ \gamma_2 \text{gdp\_per\_capita}^t \\
&\quad + \gamma_3 \text{Economic Sophistication Proxy}^t \\
&\quad + \delta_1 \text{daynight} 
+ \delta_2 \text{is\_summer} \\
&\quad + \mathbf{\theta} \cdot \text{Region Fixed Effects} 
+ \epsilon^t,
\end{aligned}
\end{equation}

The dependent variable, Political Polarization, is measured by three entropy values for each year. These include the \textit{2018 Presidential Entropy}, which is the unweighted entropy value, and two weighted versions: one weighted by the religious inclination degree and another weighted by the political inclination degree of candidates and parties. Similarly, the entropy values for 2019 include the unweighted and the two weighted measures. These entropy measures serve as comprehensive indicators of political polarization, capturing both the diversity and the degree of alignment among voter preferences.

The RHS variables include a set of economic and social controls, as well as fixed effects for regions.
Region Fixed Effects ($\mathbf{\theta} \cdot \text{Region}$): Provinces are stratified into regions based on geographic location: \textit{central, east, marmara, south, southeast, west}. Stratification addresses the skewed distribution of data across provinces, as smaller provinces tend to lack sufficient observations while larger provinces are overrepresented. By grouping provinces into regions, this fixed effect controls for spatial variations in political polarization.

Together, these controls and fixed effects provide a comprehensive framework to isolate the determinants of political polarization, considering both temporal and spatial dimensions.

\begin{table}[ht]
\centering
\caption{Regression Results for 2018 Political Entropy}
\label{tab:2018_regression_results}
\resizebox{\textwidth}{!}{%
\begin{tabular}{lccc}
\toprule
\textbf{Variable} & \textbf{Unweighted Entropy} & \textbf{Religious Weighted Entropy} & \textbf{Political Weighted Entropy} \\
\midrule
\textbf{Constant}                   & 1.1155*** (0.009) & 0.5298*** (0.004) & 0.6095*** (0.005) \\
\textbf{NRP vs RP}                  & 0.0055 (0.006)    & 0.0009 (0.003)    & 0.0046 (0.003)*   \\
\textbf{NRP vs NRP}                 & -0.0267*** (0.006) & -0.0117*** (0.002) & -0.0128*** (0.003) \\
\textbf{Number of Mosques}          & 4.51e-05*** (2.81e-06) & 1.96e-05*** (1.19e-06) & 1.88e-05*** (1.40e-06) \\
\textbf{GDP per Capita}             & 3.83e-06*** (1.66e-07) & 2.80e-06*** (7.02e-08) & 2.74e-06*** (8.27e-08) \\
\textbf{Economic Sophistication Proxy} & -2.8207*** (0.039) & -2.2828*** (0.017) & -2.3214*** (0.02) \\
\textbf{Daynight}                   & -0.0056 (0.004)   & -0.0015 (0.002)   & -0.0024 (0.002)   \\
\textbf{Is Summer}                  & -0.0103*** (0.002) & -0.0024** (0.001) & -0.0017 (0.001)   \\
\textbf{Central (Baseline)}         & 0.1097*** (0.004) & 0.0454*** (0.002) & 0.0531*** (0.003) \\
\textbf{East}                       & 0.0904*** (0.009) & 0.0794*** (0.003) & 0.0278*** (0.004) \\
\textbf{Marmara}                    & 0.1856*** (0.005) & 0.0686*** (0.003) & 0.0796*** (0.003) \\
\textbf{South}                      & 0.3713*** (0.005) & 0.1335*** (0.002) & 0.152*** (0.002)  \\
\textbf{Southeast}                  & 0.0490*** (0.006) & 0.0188*** (0.002) & -0.0174*** (0.003) \\
\textbf{West}                       & 0.0956*** (0.004) & 0.0275*** (0.002) & 0.0218*** (0.002) \\
\midrule
Observations                        & 14,683            & 14,683            & 14,683            \\
\( R^2 \)                            & 0.723             & 0.914             & 0.889             \\
Adjusted \( R^2 \)                   & 0.723             & 0.914             & 0.889             \\
F-statistic                         & 2944.0***         & 12034.0***        & 9025.0***         \\
\bottomrule
\end{tabular}%
}
\end{table}

\begin{table}[ht]
    \centering
    \caption{Regression Results for 2019 Political Polarization}
    \resizebox{\textwidth}{!}{%
    \begin{tabular}{lccc}
        \hline
        & \textbf{Unweighted} & \textbf{Religious Weighted} & \textbf{Political Weighted} \\
        \hline
        \textbf{const} & 1.8366 (0.012)*** & 0.9741 (0.005)*** & 0.8531 (0.005)*** \\
        \textbf{NRP\_vs\_RP} & 0.0510 (0.008)*** & 0.0126 (0.003)*** & 0.0157 (0.003)*** \\
        \textbf{NRP\_vs\_NRP} & -0.0291 (0.008)*** & -0.0105 (0.003)*** & -0.0151 (0.003)*** \\
        \textbf{Num\_Mosques} & 0.0001 (3.67e-06)*** & 2.804e-05 (1.43e-06)*** & -1.515e-05 (1.53e-06)*** \\
        \textbf{GDP\_per\_Capita} & -5.189e-06 (1.91e-07)*** & -3.531e-06 (7.48e-08)*** & 7.002e-07 (7.97e-08)*** \\
        \textbf{Economic Sophistication Proxy} & 0.5211 (0.055)*** & 0.2563 (0.022)*** & 0.4655 (0.023)*** \\
        \textbf{Daynight} & -0.0150 (0.005)** & -0.0064 (0.002)*** & -0.0067 (0.002)*** \\
        \textbf{Is\_summer} & -0.0016 (0.005) & -0.0016 (0.001) & 0.0010 (0.001) \\
        \textbf{Central} & 0.0593 (0.006)*** & 0.0136 (0.002)*** & 0.0503 (0.003)*** \\
        \textbf{East} & -0.1970 (0.011)*** & -0.0607 (0.004)*** & 0.0299 (0.005)*** \\
        \textbf{Marmara} & -0.0072 (0.003) & 0.0151 (0.003)*** & 0.0082 (0.003) \\
        \textbf{South} & 0.3430 (0.006)*** & 0.0828 (0.002)*** & 0.1773 (0.003)*** \\
        \textbf{Southeast} & -0.1024 (0.003)*** & -0.0637 (0.003)*** & 0.0243 (0.003)*** \\
        \textbf{West} & 0.0913 (0.006)*** & 0.0243 (0.002)*** & 0.0871 (0.002)*** \\
        \hline
        \textbf{R-squared} & 0.339 & 0.361 & 0.412 \\
        \textbf{Adj. R-squared} & 0.339 & 0.361 & 0.411 \\
        \textbf{F-statistic} & 579.5 & 638.2 & 789.4 \\
        \hline
\end{tabular}%
}
\end{table}

\section{Discussion of Results}

The regression results reveal several important insights into the relationship between sociocultural distances and electoral polarization in Turkey. By carefully structuring the model with a constant term and omitting \textit{RP vs RP} as the reference category, the coefficients for \textit{NRP vs RP} and \textit{NRP vs NRP} provide interpretable measures of how these distances deviate from the baseline represented by the constant term.

\subsection{Key Findings and Economic Interpretation}

\subsubsection{Role of \textit{NRP vs RP}}
Across all models, the coefficient for \textit{NRP vs RP} generally shows a positive and significant relationship (at least in 2019 for weighted entropies), suggesting that larger sociocultural distances between non-religious and religious individuals contribute to higher levels of electoral entropy. This aligns with theoretical frameworks such as Alesina et al.'s \textit{Fractionalization}, which emphasize that intergroup differences often exacerbate societal fragmentation and polarization \cite{alesina2003fractionalization}.

The economic implication is that intergroup frictions—such as those between secular and religious groups—manifest in political outcomes as higher entropy, reflecting diversity or division in electoral choices.

\subsubsection{Role of \textit{NRP vs NRP}}
The coefficient for \textit{NRP vs NRP} is consistently negative and significant, indicating that within-group heterogeneity among non-religious individuals has a stabilizing effect on political entropy relative to the baseline (\textit{RP vs RP}). This suggests that internal diversity within secular groups may dilute the polarizing effects of intergroup interactions.

This finding can be understood in light of Axelrod’s \textit{Dissemination of Culture}\cite{Axelrod1997}, where internal heterogeneity within a group can dampen the clustering effects of cultural divisions.

\subsubsection{Baseline and the Role of the Constant Term (\textit{RP vs RP})}
The constant term represents the baseline electoral entropy when distances between religious individuals (\textit{RP vs RP}) are the dominant factor. The relatively large magnitude of the constant across all models suggests that within-group cohesion among religious individuals inherently contributes to electoral entropy, albeit less dynamically than intergroup distances.

\subsection{Control Variables}
\subsubsection{Economic Sophistication and GDP}
The proxies for economic development, including \textit{Economic Sophistication Proxy} and \textit{GDP per capita}, show significant effects on entropy. Higher economic sophistication is associated with increased entropy, possibly due to the diversifying effects of industrialization and urbanization on voter preferences.

\subsubsection{Religiosity Proxy (\textit{num\_mosques})}
The number of mosques, serving as a proxy for religiosity, has a consistently positive and significant effect on entropy. This suggests that higher religiosity within a province correlates with greater electoral diversity, reflecting the complex interplay of religious identity and political behavior.

\subsubsection{Temporal and Regional Factors}
The time and region fixed effects capture unobserved heterogeneity across years and geographic regions, respectively. Significant coefficients for regional dummies highlight the spatial variation in political dynamics, while the inclusion of time-fixed effects ensures that changes across 2018 and 2019 are adequately controlled.

\subsection{Environmental Factors}
\textit{Is\_summer} has a significant coefficient, it indicates that seasonal factors systematically affect walking distances, highlighting the need to control for this variation to isolate the effects of sociocultural or political factors.

In summer, outdoor public spaces might be more crowded due to favorable weather, leisure activities, or tourism. This increased activity could reduce average walking distances due to practical constraints like crowded streets or public events.

Conversely, higher temperatures may discourage individuals from lingering in public spaces or influence their route choices to avoid direct sunlight or heat, potentially affecting walking patterns.
\subsection{Broader Implications}
These findings contribute to the broader literature on political polarization by providing empirical support for the role of sociocultural distances in shaping electoral outcomes. The results align with Alesina’s work on ethnic divisions and Barro’s analyses of religion, suggesting that sociocultural fragmentation—whether along religious or secular lines—is a critical driver of political entropy.

The results also demonstrate that the methodological approach—quantifying distances between groups based on objective sociocultural metrics—is a promising avenue for studying polarization. The scalability and replicability of this approach make it applicable to other contexts, such as ethnoreligious divisions in Indonesia or ideological divides in the United States.

\section{Multicollinearity and Omitted Variable Bias Analysis}

To ensure the robustness of the regression results, we conducted a multicollinearity check using the Variance Inflation Factor (VIF) and evaluated the potential for omitted variable bias (OVB).

\subsection{Multicollinearity Check}

Multicollinearity arises when independent variables are highly correlated, potentially distorting coefficient estimates and inflating standard errors. Table~\ref{tab:vif} presents the VIF values for each variable.

\begin{table}[h!]
\centering
\caption{Variance Inflation Factor (VIF) Analysis}
\label{tab:vif}
\begin{tabular}{|l|c|}
\hline
\textbf{Variable} & \textbf{VIF} \\
\hline
NRP\_vs\_RP & 1.05 \\
NRP\_vs\_NRP & 1.03 \\
num\_mosques & 4.40 \\
gdp\_per\_capita & 8.33 \\
Economic Sophistication Proxy & 67.60 \\
daynight & 1.01 \\
is\_summer & 1.02 \\
region fix effects & 56.23\\
\hline
\end{tabular}
\end{table}

Most VIF values are below the commonly accepted threshold of 10, indicating no immediate concern for multicollinearity. However, the \textit{Economic Sophistication Proxy} and \textit{region fix effect dummies} exhibits a notably high VIF, likely due to its correlation with GDP per capita or directly with LHS, which is the voting outcome. While theoretically justified, this necessitates caution when interpreting its coefficient. The unusual high $R^2$ could be influenced by the dominant predictability of region fix effects and economic status. This may imply that Turkiye is a highly divided country, not just politically, but demographically and economically. The geographic location almost indicates the features of the province. 

\subsection{Omitted Variable Bias Analysis - auxiliary regression }

OVB occurs when relevant variables are excluded from the model, biasing the estimated coefficients of included variables. An auxilliary regression is run to examine the OVB.

The auxiliary regression introduces \textit{poverty rate}, calculated as the proportion of households below 60\% of the median income, as a control variable to examine its influence on the entropy measures. This variable provides a proxy for regional wealth inequality, enabling an assessment of how economic conditions relate to polarization metrics.

A negative coefficient is observed for all entropy measures. For example, in the local entropy regression, a unit increase in the poverty rate is associated with a \(-1.341*10exp(-0.5)\) reduction in entropy, which is statistically significant. This suggests that higher poverty rates are associated with less political diversity or polarization, potentially indicating that economically disadvantaged regions may have more homogenous voting behaviors or more prone to a dominant party's propaganda. This aligns with the finding for other economic controls - richer and more urban regions indicate diverse political voting patterns while poorer and rural regions indicate homogenous inclinations.

\begin{align*}
\text{Political Polarization}_i^t &= \beta_0 
+ \beta_1 \text{NRP\_vs\_RP}_i^t 
+ \beta_2 \text{NRP\_vs\_NRP}_i^t \\
&\quad + \gamma_1 \text{num\_mosques}_i^t 
+ \gamma_2 \text{gdp\_per\_capita}_i^t \\
&\quad + \gamma_3 \text{Economic\_Sophistication\_Proxy}_i^t 
+ \delta_1 \text{daynight}_i 
+ \delta_2 \text{is\_summer}_i \\
&\quad + \eta \text{poverty\_rate}_i^t 
+ \mathbf{\theta} \cdot \text{Region Fixed Effects}_i 
+ \epsilon_i^t,
\end{align*}

The auxiliary regressions have modest explanatory power, with R-squared values ranging from 0.088 to 0.252. While poverty rate improves model fit slightly, much of the variation remains unexplained, highlighting potential omitted variable bias. Factors like urban density, infrastructure, and cultural norms could also influence the results but are not accounted for in the current model. Moreover, the use of aggregated sub-regional poverty data introduces imprecision, as finer provincial-level data would likely enhance robustness.

The negative association between poverty rate and polarization may reflect a homogenizing effect of poverty in limiting diverse political preferences. Economically disadvantaged groups might prioritize material needs over ideological divides, leading to less fragmented voting patterns. However, these interpretations should be approached cautiously, acknowledging the limitations of the lacking provincial data and the possibility of multicollinearity due to over-control of the regional fix effects (demographic and economic controls are bijective).

\begin{table}[H]
\centering
\caption{Regression Results for 2018 Entropy Variables with Poverty Rate}
\label{tab:2018_aux_regression}
\resizebox{\textwidth}{!}{%
\begin{tabular}{lccc}
\hline
 & \textbf{2018 Presidential Entropy} & \textbf{2018 Religious Entropy} & \textbf{2018 Political Entropy} \\
\hline
\textbf{Constant} & 1.2728*** (0.012) & 0.5990*** (0.005) & 0.6265*** (0.006) \\
\textbf{NRP vs RP} & 0.0207*** (0.008) & 0.0063** (0.003) & 0.0131*** (0.004) \\
\textbf{NRP vs NRP} & -0.0357*** (0.007) & -0.0149*** (0.003) & -0.0156*** (0.003) \\
\textbf{Number of Mosques} & 8.718e-06*** (2.93e-06) & 7.207e-06*** (1.19e-06) & 8.163e-06*** (1.41e-06) \\
\textbf{GDP per Capita 2018} & 1.155e-05*** (2.62e-07) & 5.586e-06*** (1.07e-07) & 5.973e-06*** (1.26e-07) \\
\textbf{Economic Sophistication Proxy 2018} & -3.3808*** (0.037) & -2.4680*** (0.015) & -2.6008*** (0.018) \\
\textbf{Day-Night Indicator} & -0.0020 (0.004) & 0.0001 (0.002) & -0.0012 (0.002) \\
\textbf{Is Summer} & -0.0023 (0.003) & -0.0071*** (0.001) & -0.0069*** (0.001) \\
\textbf{Poverty Rate 2018} & -2.336e-05*** (1.36e-06) & -9.619e-06*** (5.55e-07) & -6.771e-06*** (6.57e-07) \\
\hline
\textbf{Observations} & 14,683 & 14,683 & 14,683 \\
\textbf{R-Squared} & 0.589 & 0.883 & 0.846 \\
\textbf{Adj. R-Squared} & 0.589 & 0.883 & 0.846 \\
\textbf{F-Statistic} & 2,627.0*** & 13,790.0*** & 10,070.0*** \\
\hline
\multicolumn{4}{l}{\footnotesize *** p$<$0.01, ** p$<$0.05, * p$<$0.1. Standard errors are in parentheses.}
\end{tabular}%
}
\end{table}

\begin{table}[ht]
\centering
\caption{Regression Results for 2019 Local Entropy, Religious Entropy, and Political Entropy (Including Poverty Rate as Auxiliary Variable)}
\label{tab:auxiliary_regression_2019}
\resizebox{\textwidth}{!}{%
\begin{tabular}{lccc}
\hline
\textbf{Variable} & \textbf{Local Entropy} & \textbf{Religious Entropy} & \textbf{Political Entropy} \\
\hline
Constant & 1.8012 (0.015)*** & 0.9312 (0.005)*** & 0.9115 (0.006)*** \\
NRP\_vs\_RP & 0.0586 (0.010)*** & 0.0158 (0.003)*** & 0.0153 (0.004)*** \\
NRP\_vs\_NRP & -0.0490 (0.009)*** & -0.0137 (0.003)*** & -0.0271 (0.004)*** \\
Num\_Mosques & 9.619e-05 (3.68e-06)*** & 2.218e-05 (1.32e-06)*** & -1.688e-05 (1.59e-06)*** \\
GDP\_Per\_Capita\_2019 & 2.013e-06 (2.71e-07)*** & -1.7e-06 (9.77e-08)*** & 3.499e-06 (1.17e-07)*** \\
Economic\_Sophistication\_Proxy\_2019 & -1.2496 (0.049)*** & -0.1155 (0.018)*** & -0.3257 (0.021)*** \\
Daynight & -0.0085 (0.006) & -0.0052 (0.002)** & -0.0020 (0.002) \\
Is\_Summer & -0.0108 (0.003)*** & -0.0047 (0.001)*** & -0.0052 (0.001)*** \\
Poverty\_Rate\_2019 & -1.341e-05 (1.31e-06)*** & -5.231e-07 (4.7e-07) & -7.672e-06 (5.64e-07)*** \\
\hline
Observations & 14,683 & 14,683 & 14,683 \\
R-squared & 0.088 & 0.252 & 0.125 \\
Adj. R-squared & 0.087 & 0.251 & 0.125 \\
F-statistic & 176.1 & 616.7 & 262.7 \\
\hline
\multicolumn{4}{l}{\footnotesize{*** $p < 0.01$, ** $p < 0.05$, * $p < 0.1$}}
\end{tabular}%
}
\end{table}

Despite these measures, certain limitations persist:
1. Unobserved Sociocultural Factors: Variables like historical political affiliations and education levels were not included, which could influence sociocultural distances and electoral entropy. 2. Endogeneity Concerns: Sociocultural distances (\textit{NRP\_vs\_RP} and \textit{NRP\_vs\_NRP}) might be influenced by factors also affecting entropy, potentially introducing bias.

\subsection{Interpretation of Multicollinearity and Overestimation Risk}

Turkiye is a highly divided country, where voting behavior is shaped by regional divisions and disparities in economic development. The significant multicollinearity observed—particularly between \textit{Economic Sophistication Proxy} and GDP per capita—reflects this structural diversity. However, such multicollinearity can inflate standard errors and potentially overestimate the coefficients of highly correlated variables. Additionally, the limited population influx between provinces leads to highly predictable voting patterns, potentially amplifying model precision and overestimating the explanatory power of certain variables.

\subsection{Cautionary Note}

The results should be interpreted with care. While the model captures key factors influencing political polarization, the high R-squared values and significant coefficients may reflect structural predictability in Turkiye’s political and social context rather than universal dynamics. The inclusion of a constant term in place of the omitted \textit{RP\_vs\_RP} interaction ensures proper reference category specification but does not fully resolve the challenge of perfect collinearity in sociocultural distances.

By addressing multicollinearity and OVB concerns, the analysis provides a robust yet cautious interpretation of the relationship between sociocultural distances and political polarization.

\section{Limitations}
This study contributes to the burgeoning intersection of computational social science and economics by introducing a novel proxy for political polarization. The approach leverages walking distances, derived from street-level imagery, to provide insights into societal divisions. While this method offers unique advantages, the study acknowledges several limitations that warrant further investigation and caution in interpretation.

\subsection{Benchmarking and Future Research}
One notable limitation is the inability to benchmark the proposed measure against traditional proxies for political polarization, such as survey data or provincial-level socio-economic indicators. Due to data unavailability at the provincial level, this comparison remains speculative. Future research could address this by exploring alternative datasets or conducting surveys tailored to polarization metrics.

\subsection{Behavioral Complexity of Walking Distance}
Walking distance, while computationally feasible, may not fully encapsulate the nuanced intentions behind individuals' walking behavior. Subtle factors such as daily moods, social contexts, or transient environmental stimuli might influence walking patterns, adding layers of variability that this study cannot account for. Additionally, the street imagery approach treats pedestrians as isolated individuals, overlooking potential group dynamics or inter-individual correlations that may alter walking distances.

\subsection{Feasibility Constraints on Behavioral Measures}
Although walking distance is computationally tractable, other behavioral interactions, such as nodding, facial expressions, or body language, could provide richer insights into fractionalization and dissemination. However, these measures demand high computational resources and large, annotated training datasets, which are currently beyond the scope of this study. Future advancements in computer vision might make such analyses more accessible.

\subsection{Religiosity Measurement Challenges}
A further limitation lies in the lack of clear and reliable measures of religiosity at the provincial level in Turkey. While the country's official records indicate that 99.5\% of the population identifies as Muslim, this statistic does not reflect the diversity of religious practice or belief. Many individuals are indifferent or even critical of Muslim norms and political influences, complicating the interpretation of religiosity's impact on polarization. Without accurate provincial-level religiosity data, the analysis remains constrained in assessing the interplay between religiosity and political division.

\subsection{Replicability To Non-religiously Divided Nations}
This study applies its insights to Turkey, a nation marked by ideological divides between secularism and political Islam. Turkey's public walking patterns provide a unique lens to uncover subconscious biases and cultural frictions, particularly between non-religious males and religious females. While these findings are relevant to Turkey, the replicability of this method in other contexts varies significantly based on the visibility of cultural or religious markers in public spaces.

In countries like Iran\cite{shahidian2002women} (Caliskan's lab is embarking on) or Indonesia \cite{fealy2008expressing}, where Islamic dress codes are strictly enforced, extending this research would be relatively straightforward. Public walking patterns could similarly reflect sociocultural divisions between groups with visible attire-based distinctions. For instance, the divide between those adhering to strict religious dress codes and those resisting such norms provides a comparable setting for applying the same analytical framework.

However, replicability becomes more challenging in countries where cultural or ideological divisions are not visibly marked by attire. Japan, for example, poses difficulties as public behavior is shaped more by norms of politeness and personal space than by outward displays of cultural or religious identity. In such contexts, walking patterns may be less indicative of polarization, necessitating alternative measurement strategies. Researchers might need to incorporate complementary data sources, such as surveys or analyses of verbal and non-verbal communication patterns, to capture polarization dynamics more effectively.

\section{Conclusion}
This study introduces a novel methodological approach to understanding political polarization by leveraging computer-vision extracted walking distances as a proxy for sociocultural divisions. The regression results provide robust empirical evidence for the role of sociocultural distances—particularly between secular and religious groups—in shaping electoral polarization, as captured through entropy measures.

Key findings reveal that larger intergroup distances (NRP vs RP) between non-religious and religious individuals significantly increase electoral entropy, underscoring the fragmenting effects of sociocultural divisions. In contrast, within-group heterogeneity among secular individuals (NRP vs NRP) has a stabilizing effect, mitigating polarization and aligning with theoretical insights from Axelrod’s cultural dissemination framework.

From an economic perspective, this result advances the understanding of political polarization by integrating sociocultural metrics with conventional economic indicators. Variables such as economic sophistication and GDP per capita further illustrate the diversifying effects of modernization, while proxies for religiosity (e.g., number of mosques) reinforce the link between religious identity and political outcomes.

From computational social science perspective, this result demonstrates the value of state-of-art computer vision techniques for quantifying sociocultural dynamics. By addressing limitations in traditional survey-based methods and endogeneity-prone economic indicators, walking distances provide an objective, scalable, and replicable measure of social interaction. This approach represents a methodological advancement with broader applicability to contexts characterized by sociocultural fragmentation, such as ethno-religious divides or ideological polarization in other regions.

Future research can build on this study by benchmarking walking distances against traditional polarization proxies, such as survey data or localized socio-economic indicators, once they become available. Additionally, advancements in computer vision and behavioral analysis—such as incorporating group dynamics, facial expressions, or other interactions—could provide richer insights into societal divisions. These refinements would enhance the robustness and generalizability of walking distances as a proxy for political polarization across diverse socio-political contexts.

Future research can also explore how walking distance analyses, combined with computational techniques, can be cross-examined with existing economic frameworks to understand polarization in different nations. In contexts like Iran or Indonesia, where visible religiosity plays a significant role, researchers could extract public interaction data from platforms such as YouTube to study behavioral patterns and their links to socio-political divides. This approach would allow for easy replication of Türkiye’s methodology, leveraging openly available digital content to identify polarization markers. In contrast, for nations like Japan, where visible religiosity is less pronounced, economists could shift focus to other polarization symbols, such as urban versus regional dialects in public speech, using Large Language Models (LLM) such as Transformer\cite{vaswani2017attention}. 

\section{Acknowledgement}
I extend my heartfelt gratitude to Michael Rizzio for his invaluable support throughout this semester as I embarked on writing my first economics research paper. His patience and encouraging words have profoundly influenced me and made this journey both enriching and inspiring.

I would also like to thank Cantay Caliskan, my data science professor, research mentor for two years, and dearest friend. His guidance, especially in our discussions about Türkiye and data mining, has been instrumental in shaping my academic and professional growth.

Special thanks to Edward Huang, my classmate and friend, whose weekly insights during our coffee chats have been a constant source of inspiration and learning.

Thanks to Kathy Wu for her dedicated, hands-on guidance in preparing the control datasets and pushing the boundaries of implementing these controls effectively.

Gratitude also goes to Prof. Katherine Schaefer for her review of my paper, her keen ability to identify areas of obscurity in my writing, and her clear, actionable advice for improving clarity and coherence.

Lastly, I express my deepest appreciation to Dr. Bryce Dietrich, a name that stands prominently among top computational social scientists in the United States. Meeting him this summer and receiving his guidance on two of my papers has left an indelible mark on me. I greatly admire his work and aspire to follow in his footsteps as I continue my journey in computational social science.

\appendix
\section*{Appendix A: Technical Details of Caliskan's Paper \cite{caliskan2025polarization}}  
\begin{figure}
    \centering
    \includegraphics[width=\linewidth]{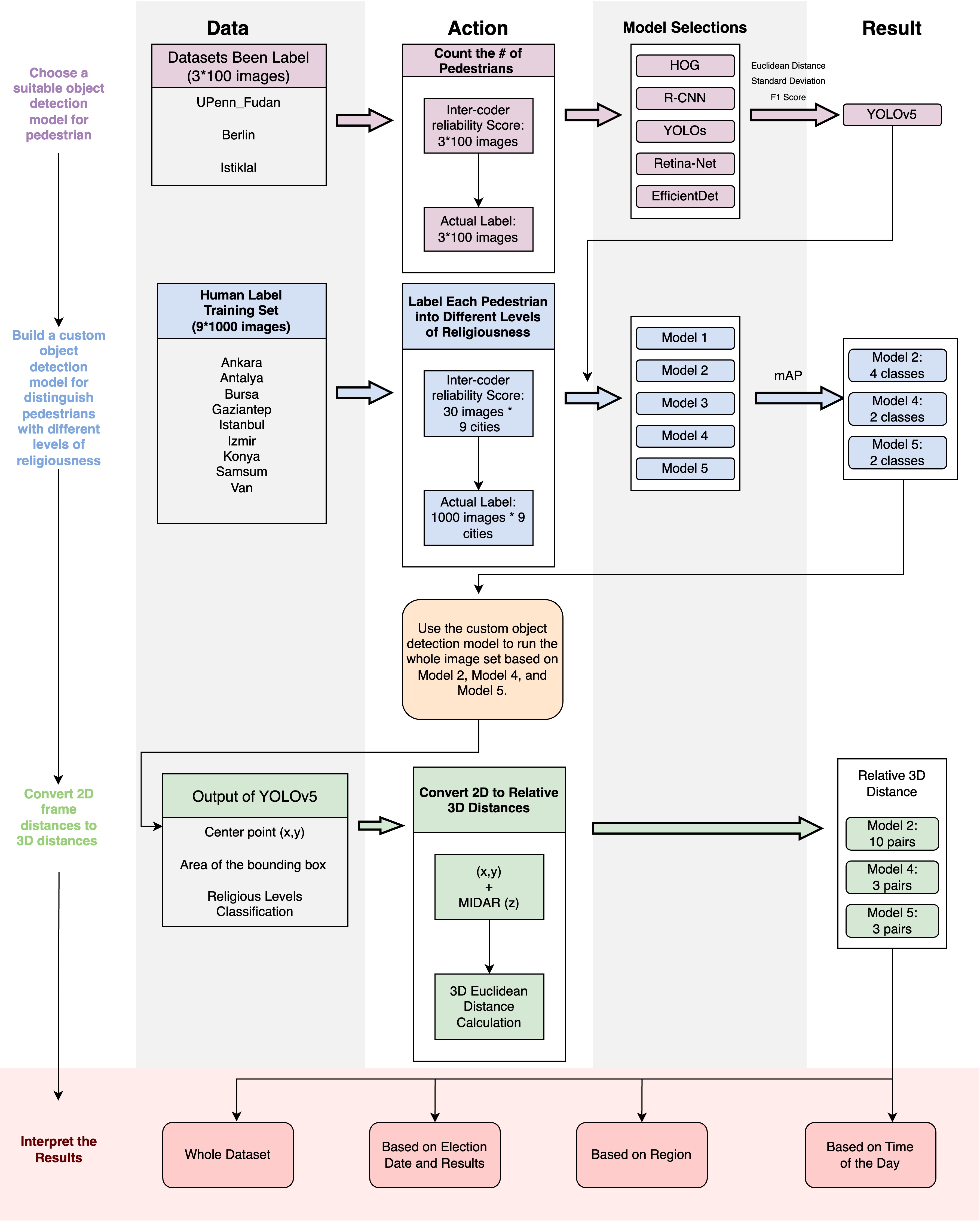}
    \caption{Pipeline For Obtaining Distance Dataset}
    \label{fig:pip}
\end{figure}

\begin{figure}[H]
    \centering
    \begin{subfigure}[b]{0.45\linewidth}
        \centering
        \includegraphics[width=\linewidth]{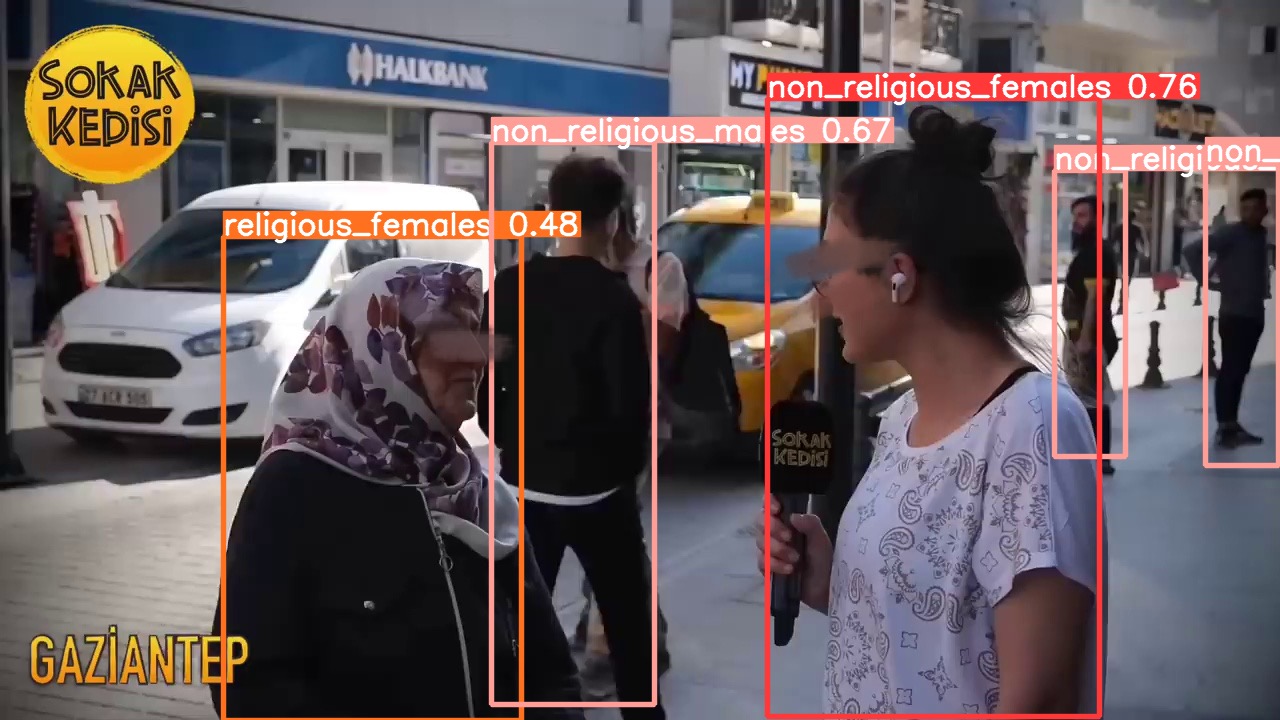}
        \caption{Enter Caption for Model 2}
        \label{fig:model2}
    \end{subfigure}
    \hfill
    \begin{subfigure}[b]{0.45\linewidth}
        \centering
        \includegraphics[width=\linewidth]{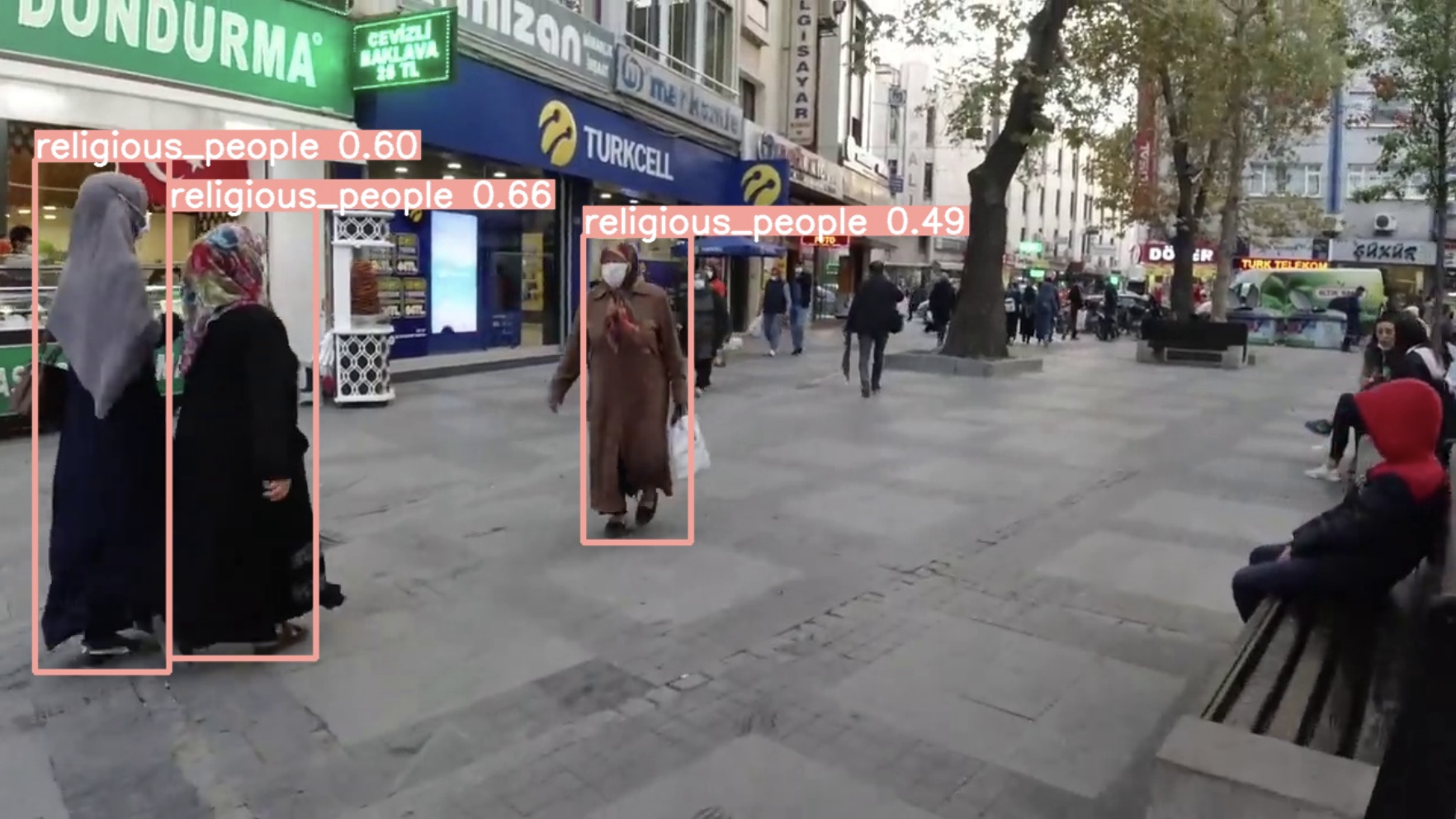}
        \caption{Enter Caption for Model 5}
        \label{fig:model5}
    \end{subfigure}
    \caption{Sample Outputs}
    \label{fig:sample-outputs}
\end{figure}

The details are summarized in Figure \ref{fig:pip}, and sample outputs are shown in Figure \ref{fig:sample-outputs}.  

\subsection*{Step 1: Choose a Suitable Object Detection Model for Pedestrians}  
\textit{Goal:} Identify the optimal object detection algorithm for detecting pedestrians.  

\begin{enumerate}  
    \item \textit{Datasets Used:} UPenn\_Fudan, Berlin, and Istiklal datasets, each with 3×100 labeled images.  

    \item \textit{Methodology:}  
    \begin{enumerate}  
        \item Count pedestrians in the datasets to ensure detection accuracy.  
        \item Calculate inter-coder reliability scores for labeling consistency.  
        \item Evaluate object detection models: HOG, R-CNN, YOLO versions, RetinaNet, and EfficientDet.  
        \item Benchmark models using Euclidean Distance, Standard Deviation, and F1 score.  
    \end{enumerate}  

    \item \textit{Result:} YOLOv5 is selected for its superior accuracy and reliability.  
\end{enumerate}  

\subsection*{Step 2: Scrape YouTube Videos and Obtain Images}  
\textit{Goal:} Build a custom object detection model to classify pedestrians by levels of religiosity.  

\begin{enumerate}  
    \item \textit{Source:} YouTube videos recorded in nine cities across Türkiye: Ankara, Antalya, Bursa, Gaziantep, Istanbul, Izmir, Konya, Samsun, and Van.  

    \item \textit{Data Collection:} Approximately 1,000 images are obtained per city.  

    \item \textit{Labeling Process:}  
    \begin{enumerate}  
        \item Pedestrians are categorized by levels of religiosity through manual annotation.  
        \item Inter-coder reliability scores are calculated for 30 images per city to ensure consistency.  
        \item The final dataset includes 9,000 labeled images across nine cities.  
    \end{enumerate}  
\end{enumerate}  

\subsection*{Step 3: Train a Custom Object Detection Model}  
\textit{Goal:} Train and fine-tune models to distinguish pedestrians based on levels of religiosity.  

\begin{enumerate}  
    \item \textit{Model Training:} YOLOv5-based models are trained using the labeled dataset with the following configurations:  
    \begin{enumerate}  
        \item Model 2: 4-class classification.  
        \item Model 4: 2-class classification (collapsed religiosity levels).  
        \item Model 5: 2-class classification with alternative labeling.  
    \end{enumerate}  

    \item \textit{Performance Evaluation:} Mean Average Precision (mAP) is used to assess classification accuracy.  
\end{enumerate}  

\subsection*{Step 4: Convert 2D Frame Distances to 3D Distances}  
\textit{Goal:} Accurately measure spatial relationships between pedestrians in 3D space.  

\begin{enumerate}  
    \item YOLOv5 outputs center point coordinates (\textit{x, y}), bounding box dimensions, and religiosity classifications.  
    \item MIDAR is used for depth estimation to derive relative \textit{z}-coordinates.  
    \item 3D Euclidean distances are calculated using (\textit{x, y, z}) coordinates.  
\end{enumerate}

\bibliographystyle{plain}  
\bibliography{cite}  

\begin{thebibliography}{10}

\bibitem{hurriyet_erdogan_nomination}
{AKP formally nominates Erdoğan for presidential election}.
\newblock {\em Hürriyet Daily News}.
\newblock Retrieved 4 May 2018.

\bibitem{hurriyet_joint_application}
{AKP, MHP jointly apply to election board for Erdoğan's presidential bid}.
\newblock {\em Hürriyet Daily News}.
\newblock Retrieved 4 May 2018.

\bibitem{turkish_minute_aksener_signatures}
{Akşener collects 100,000 signatures in 4 hours to become presidential candidate}.
\newblock {\em Turkish Minute}.
\newblock Retrieved 4 May 2018.

\bibitem{hurriyet_ince_president}
{CHP presidential candidate İnce vows to be 'everyone's president'}.
\newblock {\em Hürriyet Daily News}.
\newblock Retrieved 4 May 2018.

\bibitem{sozcuk_kesici}
{CHP'li İlhan Kesici: Adayımız Muharrem İnce}.
\newblock {\em Sözcü}.
\newblock Retrieved 3 May 2018.

\bibitem{balkan_bosnia_rally}
{Erdogan's Bosnia Rally May Be Key Game-Changer}.
\newblock {\em Balkan Insight}.
\newblock Retrieved 4 May 2018.

\bibitem{hurriyet_demirtas_nomination}
{HDP nominates imprisoned former leader Demirtaş for presidency}.
\newblock {\em Hürriyet Daily News}.
\newblock Retrieved 4 May 2018.

\bibitem{hurriyet_ince_nomination}
{Muharrem İnce likely presidential nominee of Turkey's main opposition CHP}.
\newblock {\em Hürriyet Daily News}.
\newblock Retrieved 3 May 2018.

\bibitem{OED_polarization}
Polarization.
\newblock Oxford English Dictionary.
\newblock 3rd ed., Oxford University Press, OED Online. Accessed 20 Oct. 2024.

\bibitem{karamollaoglu_socialmedia}
{Türkiye'ye Bilge Başkan! Sosyal medyada gündem Temel Karamollaoğlu}.
\newblock Retrieved 24 June 2018.

\bibitem{ackson2017islamicsecular}
S.~A. Ackson.
\newblock The islamic secular.
\newblock {\em American Journal of Islam and Society}, 34(2):1--38, 2017.

\bibitem{Alaimo2023}
Luca~S. Alaimo.
\newblock The complexity of social phenomena and the construction of indicators.
\newblock In Enrica di~Bella, Sandra Fachelli, Pedro López-Roldán, and Christian Suter, editors, {\em Measuring Gender Equality}, volume~87 of {\em Social Indicators Research Series}, pages 35--55. Springer, Cham, 2023.

\bibitem{alesina2003fractionalization}
Alberto Alesina, Arnaud Devleeschauwer, William Easterly, Sergio Kurlat, and Romain Wacziarg.
\newblock Fractionalization.
\newblock {\em Journal of Economic Growth}, 8(2):155--194, 2003.

\bibitem{Arora2022}
Swapan~Deep Arora, Guninder~Pal Singh, Anirban Chakraborty, and Moutusy Maity.
\newblock Polarization and social media: A systematic review and research agenda.
\newblock {\em Technological Forecasting and Social Change}, 183:121942, 2022.

\bibitem{Axelrod1997}
Robert Axelrod.
\newblock The dissemination of culture: A model with local convergence and global polarization.
\newblock {\em Journal of Conflict Resolution}, 41(2):203--226, 1997.

\bibitem{aydinlik_perincek_signatures}
{Aydınlık Gazetesi}.
\newblock {Baraj yıkıldı: Doğu Perinçek 100 bini aştı!}
\newblock {\em Aydınlık}.

\bibitem{Barro2001human}
Robert~J. Barro.
\newblock Human capital and growth.
\newblock {\em American Economic Review}, 91(2):12–17, May 2001.

\bibitem{baskan2006}
Filiz Başkan.
\newblock Globalization and nationalism: The nationalist action party of turkey.
\newblock {\em Nationalism and Ethnic Politics}, 12(1):83--105, 2006.

\bibitem{Bradley2022}
M.~Bradley and S.~Chauchard.
\newblock The ethnic origins of affective polarization: Statistical evidence from cross-national data.
\newblock {\em Frontiers in Political Science}, 4:920615, 2022.

\bibitem{Bragazzi2023}
Nicola~L. Bragazzi, Mirko Converti, Annalisa Crapanzano, Raffaele Zerbetto, Alessandro Siri, and Rania Khamisy-Farah.
\newblock Probing the genomic landscape of human sexuality: a critical systematic review of the literature.
\newblock {\em Frontiers in Genetics}, 14:1184758, Aug 2023.

\bibitem{Brenner2019}
Philip~S. Brenner.
\newblock Sexuality, political polarization, and survey reports of religious nonaffiliation.
\newblock {\em Politics and Religion}, 12(1):153--170, 2019.

\bibitem{butler2011do}
Daniel~M. Butler and David~E. Broockman.
\newblock Do politicians racially discriminate against constituents? a field experiment on state legislators.
\newblock {\em American Journal of Political Science}, 55(3):463--477, 2011.
\newblock Accessed 10 Dec. 2024.

\bibitem{caliskan2025polarization}
C.~Caliskan, L.~Ke, Y.~Li, et~al.
\newblock Measuring political polarization through visible interactions between religious and non-religious citizens.
\newblock {\em Scientific Reports}, 15(33795), 2025.

\bibitem{CallanderCarbajal2022}
Steven Callander and Juan~Carlos Carbajal.
\newblock Cause and effect in political polarization: A dynamic analysis.
\newblock {\em Journal of Political Economy}, 130(6):1672--1708, 2022.

\bibitem{celep2010}
Ödül Celep.
\newblock Turkey’s radical right and the kurdish issue: The mhp’s reaction to the democratic opening.
\newblock {\em Insight Turkey}, 12(2):125--142, 2010.

\bibitem{cengiz2021}
Fatih~Çağatay Cengiz.
\newblock Resistance to change: The ideological immoderation of the nationalist action party in turkey.
\newblock {\em Turkish Studies}, 22(2):462--480, 2021.

\bibitem{DietrichSands2023}
B.J. Dietrich and M.L. Sands.
\newblock Seeing racial avoidance on new york city streets.
\newblock {\em Nat Hum Behav}, 7:1275--1281, 2023.

\bibitem{DiMaggio1996}
Paul DiMaggio, John Evans, and Bethany Bryson.
\newblock Have american's social attitudes become more polarized?
\newblock {\em American Journal of Sociology}, 102(3):690--755, 1996.

\bibitem{downs1957}
Anthony Downs.
\newblock {\em An Economic Theory of Democracy}.
\newblock Harper \& Row, New York, 1957.

\bibitem{eralp_turkey_candidates_eu}
{Doğa Ulaş Eralp}.
\newblock {What do Turkey's other presidential candidates think about the EU?}
\newblock {\em Ahval}.
\newblock Archived from the original on 22 January 2021. Retrieved 31 May 2018.

\bibitem{enos2017space}
Ryan~D. Enos.
\newblock {\em The Space between Us: Social Geography and Politics}.
\newblock Cambridge University Press, 2017.

\bibitem{Evans2002}
Geoffrey Evans and Ariana Need.
\newblock Explaining ethnic polarization over attitudes towards minority rights in eastern europe: A multilevel analysis.
\newblock {\em Social Science Research}, 31(4):653--680, 2002.

\bibitem{fealy2008expressing}
Greg Fealy and Sally White, editors.
\newblock {\em Expressing Islam: Religious Life and Politics in Indonesia}.
\newblock Institute of Southeast Asian Studies, Singapore, 2008.

\bibitem{Fiorina2008}
Morris~P. Fiorina and Samuel~J. Abrams.
\newblock Political polarization in the american public.
\newblock {\em Annual Review of Political Science}, 11:563--588, 2008.
\newblock at 582.

\bibitem{Fraser2022}
Timothy Fraser, Daniel~P. Aldrich, Costas Panagopoulos, David Hummel, and Daniel Kim.
\newblock The harmful effects of partisan polarization on health.
\newblock {\em PNAS Nexus}, 1(1), March 2022.

\bibitem{Grechyna2016}
Daryna Grechyna.
\newblock On the determinants of political polarization.
\newblock {\em Economics Letters}, 144:10--14, 2016.

\bibitem{gumuscu2024}
S.~Gumuscu.
\newblock The akp and stealth islamization in turkey.
\newblock {\em Turkish Studies}, 25(3):371--397, 2024.

\bibitem{Guo2023}
Jingjing Guo and Yong Hu.
\newblock Does social media use polarize or depolarize political opinion in china? explaining opinion polarization within an extended communication mediation model.
\newblock {\em Social Media + Society}, 9(3), 2023.

\bibitem{hausmann2007complexity}
Ricardo Hausmann, Jason Hwang, and Dani Rodrik.
\newblock The product space conditions the development of nations.
\newblock {\em Science}, 317(5837):482--487, 2007.

\bibitem{Heider1946}
Fritz Heider.
\newblock Attitudes and cognitive organization.
\newblock {\em The Journal of Psychology}, 21:107--112, 1946.

\bibitem{f510ce5e7c7a4ebead2435ddb1bf808b}
{Thomas A.} Hirschl, {James G.} Booth, {Leland L.} Glenna, and {Brandn Q.} Green.
\newblock Politics, religion, and society: Is the united states experiencing a period of religious-political polarization?
\newblock {\em Review of European Studies}, 4(4):95--109, 2012.

\bibitem{Iyengar2012}
Shanto Iyengar, Gaurav Sood, and Yphtach Lelkes.
\newblock Affect, not ideology: A social identity perspective on polarization.
\newblock {\em Public Opinion Quarterly}, 76(3):405--431, November 2012.
\newblock at 428.

\bibitem{Jardina2022}
Ashley Jardina and Trent Ollerenshaw.
\newblock The polls—trends: The polarization of white racial attitudes and support for racial equality in the us.
\newblock {\em Public Opinion Quarterly}, 86(S1):576--587, 2022.

\bibitem{AKP}
Justice and Development Party.
\newblock Official website of akp, 2019.
\newblock Accessed: 2019-12-01.

\bibitem{kellam2017}
Marisa Kellam.
\newblock Why pre-electoral coalitions in presidential systems?
\newblock {\em British Journal of Political Science}, 47(2):391--411, 2017.

\bibitem{Kubin2021}
Elise Kubin and Christian von Sikorski.
\newblock The role of (social) media in political polarization: a systematic review.
\newblock {\em Annals of the International Communication Association}, 45(3):188--206, 2021.

\bibitem{Kuznets1962}
Simon Kuznets.
\newblock How to judge quality.
\newblock {\em The New Republic}.

\bibitem{Lachat2008}
Romain Lachat.
\newblock The impact of party polarization on ideological voting.
\newblock {\em Electoral Studies}, 27(4):687--698, 2008.

\bibitem{Lazer2009}
David Lazer, Alex Pentland, Lada Adamic, Sinan Aral, Albert-L{\'a}szl{\'o} Barab{\'a}si, Devon Brewer, et~al.
\newblock Computational social science.
\newblock {\em Science}, 323(5915):721--723, 2009.

\bibitem{Lee2019}
C.~Lee.
\newblock Electoral politics, party polarization, and arms control: New start in historical perspective.
\newblock {\em Orbis}, 63(4):545--564, 2019.

\bibitem{Lee2022}
C.~A. Lee.
\newblock Polarization, casualty sensitivity, and military operations: Evidence from a survey experiment.
\newblock {\em International Politics}, 59(5):981--1003, 2022.
\newblock Epub 2022 Mar 16.

\bibitem{luca2023}
Daniele Luca, Juan Terrero-Davila, Jonathan Stein, and Neil Lee.
\newblock Progressive cities: Urban–rural polarisation of social values and economic development around the world.
\newblock {\em Urban Studies}, 60(12):2329--2350, 2023.

\bibitem{McCarty2006}
Nolan McCarty, Keith~T. Poole, and Howard Rosenthal.
\newblock {\em Polarized America: The Dance of Ideology and Unequal Riches}.
\newblock MIT Press, Cambridge, MA, 2006.

\bibitem{McCoy2018}
Jennifer McCoy, Tahmina Rahman, and Murat Somer.
\newblock Polarization.
\newblock {\em Annual Review of Political Science}, 21:409--425, 2018.
\newblock at 18.

\bibitem{Meernik1993}
James Meernik.
\newblock Presidential support in congress: Conflict and consensus on foreign and defense policy.
\newblock {\em Journal of Politics}, 55(2):569--587, 1993.

\bibitem{Migheli2019}
Matteo Migheli.
\newblock Religious polarization, religious conflicts and individual financial satisfaction: Evidence from india.
\newblock {\em Review of Development Economics}, 23(2):803--829, May 2019.

\bibitem{tkp}
Communist~Party of~Turkey.
\newblock Official website of tkp, 2019.
\newblock Accessed: 2019-12-01.

\bibitem{kidal2020}
Arzu Opçin~Kıdal.
\newblock {\em Continuity and Change in the Ideology of the Nationalist Action Party (MHP), 1965-2015: From Alparslan Türkeş to Devlet Bahçeli}.
\newblock PhD thesis, Bilkent University, Ankara, 2020.

\bibitem{turkeyelections}
Ozancan Ozdemir.
\newblock {\em The Most Comprehensive R Package for Turkish Election Results}, 2024.
\newblock R package version 0.1.2.

\bibitem{dp}
Democrat Party.
\newblock Official website of dp, 2019.
\newblock Accessed: 2019-12-01.

\bibitem{dsp}
Democratic~Left Party.
\newblock Official website of dsp, 2019.
\newblock Accessed: 2019-12-01.

\bibitem{iyi}
Good Party.
\newblock Official website of iyi party, 2019.
\newblock Accessed: 2019-12-01.

\bibitem{bbp}
Great~Unity Party.
\newblock Official website of bbp, 2019.
\newblock Accessed: 2019-12-01.

\bibitem{btp}
Independent~Turkey Party.
\newblock Official website of btp, 2019.
\newblock Accessed: 2019-12-01.

\bibitem{mhp}
Nationalist~Movement Party.
\newblock Official website of mhp, 2019.
\newblock Accessed: 2019-12-01.

\bibitem{vp}
Patriotic Party.
\newblock Official website of vp, 2019.
\newblock Accessed: 2019-12-01.

\bibitem{hdp}
Peoples'~Democratic Party.
\newblock Official website of hdp, 2019.
\newblock Accessed: 2019-12-01.

\bibitem{chp}
Republican~People's Party.
\newblock Official website of chp, 2019.
\newblock Accessed: 2019-12-01.

\bibitem{saadet}
Saadet Party.
\newblock Official website of saadet party, 2019.
\newblock Accessed: 2019-12-01.

\bibitem{Reiljan2020}
Andres Reiljan.
\newblock ‘fear and loathing’: Explaining political polarization across western democracies.
\newblock {\em European Journal of Political Research}, 59(2):376--394, 2020.
\newblock at 377.

\bibitem{saadet2011}
{Saadet Partisi (Felicity Party)}.
\newblock Saadet partisi (felicity party), turkey, 2011, 2011.
\newblock Translated for the Islamic Political Party Platform Project, University of North Carolina, Chapel Hill.

\bibitem{Schedler2023}
Andreas Schedler.
\newblock Rethinking political polarization.
\newblock {\em Political Science Quarterly}, 138(3):335--359, Fall 2023.

\bibitem{secen2024}
S.~Secen, S.~Al, and B.~Arslan.
\newblock Electoral dynamics, new nationalisms, and party positions on syrian refugees in turkey.
\newblock {\em Turkish Studies}, 25(3):419--449, 2024.

\bibitem{shahidian2002women}
Hammed Shahidian.
\newblock {\em Women in Iran: Gender Politics in the Islamic Republic}.
\newblock Greenwood Publishing Group, Westport, CT, 2002.

\bibitem{shannon1948}
Claude~E. Shannon.
\newblock A mathematical theory of communication.
\newblock {\em Bell System Technical Journal}, 27:379--423, 1948.

\bibitem{slininger2014}
Shanna Slininger.
\newblock Veiled women: Hijab, religion, and cultural practice.
\newblock 2014.

\bibitem{Stanley2020}
Matthew~L. Stanley, Paul Henne, Brenda~W. Yang, and Felipe De~Brigard.
\newblock Resistance to position change, motivated reasoning, and polarization.
\newblock {\em Political Behavior}, 42(3):891--913, September 2020.
\newblock at 895.

\bibitem{tavits2008}
Margit Tavits.
\newblock Party systems in the making: The emergence and success of new parties in new democracies.
\newblock {\em British Journal of Political Science}, 38(1):113--133, 2008.

\bibitem{turkstat}
Turkish Statistical~Institute (TurkStat).
\newblock Statistics by theme.
\newblock \url{https://www.tuik.gov.tr}.
\newblock Accessed: [Insert Date].

\bibitem{us_religious_freedom_turkey_2021}
{U.S. Department of State}.
\newblock 2021 report on international religious freedom: Turkey, 2021.
\newblock Available at: \url{https://www.state.gov/reports/2021-report-on-international-religious-freedom/turkey/}.

\bibitem{Vasist2023}
P.~N. Vasist, D.~Chatterjee, and S.~Krishnan.
\newblock The polarizing impact of political disinformation and hate speech: A cross-country configural narrative.
\newblock {\em Information Systems Frontiers}, pages 1--26, April 17 2023.
\newblock Epub ahead of print.

\bibitem{vaswani2017attention}
Ashish Vaswani, Noam Shazeer, Niki Parmar, Jakob Uszkoreit, Llion Jones, Aidan~N Gomez, {\L}ukasz Kaiser, and Illia Polosukhin.
\newblock Attention is all you need.
\newblock In {\em Advances in Neural Information Processing Systems}, volume~30, 2017.

\bibitem{WilkinsLaflamme2024}
Sarah Wilkins-Laflamme, David Voas, and Kirstie Hewlett.
\newblock Religious polarization in europe.
\newblock {\em Sociology of Religion}, 2024.
\newblock srae017, Epub ahead of print.

\bibitem{yesilada2023}
B.~A. Yeşilada.
\newblock The akp, religion, and political values in contemporary turkey: Implications for the future of democracy.
\newblock {\em Turkish Studies}, 24(3--4):593--616, 2023.

\bibitem{carkoglu2007}
Ali Çarkoğlu, Binnaz Toprak, and Charlie~Andre Fromm.
\newblock {\em Religion, Society and Politics in a Changing Turkey}.
\newblock TESEV Publications, 2007.

\end{thebibliography}

\end{document}